\documentclass{iopjournal}
\expandafter\let\csname equation*\endcsname\relax
\expandafter\let\csname endequation*\endcsname\relax

\usepackage{amsmath}
\usepackage{bm}
\usepackage{graphicx}
\usepackage{amssymb}
\usepackage{color,soul}
\usepackage[T1]{fontenc}
\usepackage{ae,aecompl}
\usepackage{cases}

\usepackage{pdfpages}

\begin{document}

\articletype{Paper}

\title[Minimizing energy dissipation during programming of resistive switching memory devices using their dynamical attractor states]

\author{Valeriy~A.~Slipko$^1$, Alon~Ascoli$^2$, Fernando~Corinto$^2$ and Yuriy~V.~Pershin$^{3,*}$}

\affil{$^1$Institute of Physics, Opole University, Opole 45-052, Poland}

\affil{$^2$Department of Electronics and Telecommunications
Politecnico di Torino, Turin, Italy}

\affil{$^3$Department of Physics and Astronomy, University of South Carolina, Columbia, SC 29208 USA}

\affil{$^*$Author to whom any correspondence should be addressed.}

\email{vslipko@uni.opole.pl,alon.ascoli@polito.it, fernando.corinto@polito.it, and pershin@physics.sc.edu}


\begin{abstract}
Under certain conditions, applying a sequence of voltage pulses of alternating polarities across a resistive switching memory device induces a finite number of fixed-point attractors \textcolor{black}{in its time-averaged dynamics}, known as dynamical attractors. 
Remarkably, dynamical attractors can 
be used to program analog values into the device state without supervision. Because different pulse sequences can produce the same trajectory solution for the state in the phase space, there is strong potential for optimization, particularly regarding the energy cost of the programming phase, which this study addresses. The proposed theory-based energy minimization strategy is applied to the voltage threshold adaptive memristor (VTEAM) model, which is known for its predictive capability and adaptability in fitting a large number of resistive switching memory devices. 
The optimization design crafts ad-hoc pulse sequences that minimize the energy required to program the device into a desired dynamical attractor. The theoretical approach is also extended to cover situations where a fast programming scheme should be adopted to serve time-critical electronics applications.    
\end{abstract}


\section{Introduction} \label{sec:Intro}

The precise programming of resistive switching memory devices~\cite{ielmini_waser_2016} is crucial for various applications that depend on memristive states. These applications include analog circuits using memristive devices~\cite{5405039}, memristive neural networks~\cite{pershin2010experimental,9618724,ascoli2020}, and crossbar-based memory arrays that can also function as computing engines, \textcolor{black}{e.g. for accelerating} matrix-vector multiplications~\cite{Strachan18a,Teuscher17a,Amirsoleimani20a}. On one hand, the effectiveness of memristive technologies heavily relies on the precision during the state programming process. On the other hand, developing methods to reduce power dissipation in a memristive device during programming is critically important, especially in mobile technical systems with limited energy supplies.

In general, applying a sequence of voltage pulses of alternating polarity across a resistive switching memory device can
induce a finite number of dynamical attractors in the device's phase space, opening up novel and intriguing opportunities for device programming. 
This concept was introduced in \cite{Pershin_2019} and further explored in \cite{pershin_2019_bif_anal_TaO_mod},\cite{Slipko19a},\cite{asc_front_electr_2023}, 
and \cite{schmitt2024theoretico}. 
Moreover, in~\cite{messaris2023}, high-frequency periodic signals were used to induce controlled resistive switching transitions in non-volatile memristor devices endowed with fading memory \cite{ascoli2016}. 
Importantly, in recent years, several studies explored the theoretical basis behind the synthesis of low-power programming schemes for memristive devices~\cite{slipko2024reduction,astin2025low,slipko2025low}. 
One notable finding is that the stimulus, minimizing  Joule losses in ideal memristors, 
corresponds to constant power dissipation~\cite{slipko2024reduction}.

\begin{figure}
    \centering
    (a)\includegraphics[width=0.35\textwidth]{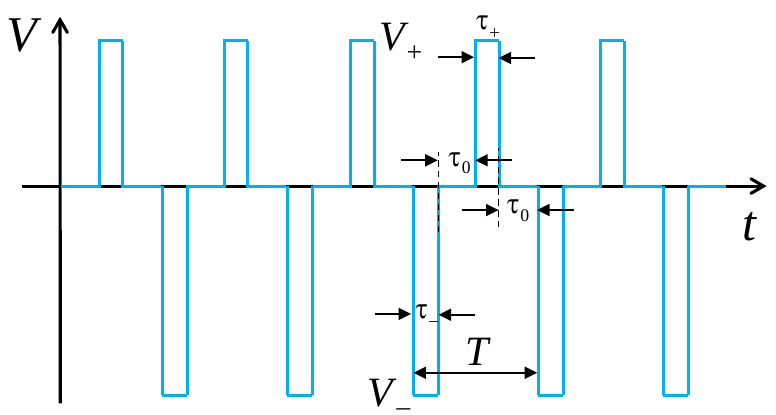}\hspace{0.5cm}
    (b)\includegraphics[width=0.35\textwidth]{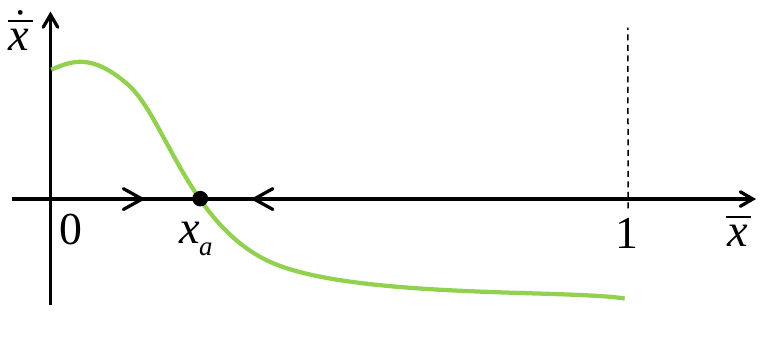}
    \caption{Illustration of the dynamical attractor concept for a first-order memristive device. 
    (a) Pulse sequence,  consisting, over each cycle, of a pair of pulses of alternating amplitudes $V_+$ and $V_-$ and corresponding widths $\tau_+$ and $\tau_-$, and let fall across the memristive device during the programming phase. 
    $T$ is the pulse train period, $\tau_0$ the inter-pulse interval. 
    (b) Schematic plot demonstrating the time-averaged flow of a time-averaged internal state variable, referred to as $\bar{x}$, for the device under the periodic pulse train excitation sketched in (a). 
    The arrow pointing to the right (left) indicates a progressive increase (decrease) in the time-averaged state from cycle to cycle. The dot marks the position of an attractor, which the pulse train stimulus induces in the phase space of the memristive device. Adapted from~\cite{pershin_2019_bif_anal_TaO_mod}.}
    \label{fig:1}
\end{figure}

This study merges the areas of dynamical attractors and low-power programming of memristive devices. Specifically, we present a theoretical framework that designs a programming pulse sequence to minimize Joule losses within the device while simultaneously producing a desired dynamic attractor \textcolor{black}{trajectory}.
Our analysis is based on the voltage threshold adaptive
memristor (VTEAM) model~\cite{Kvatinsky2013} that is a particular realization
of first-order memristive devices and systems~\cite{chua76a} 
\begin{eqnarray}
I(t)&=&G_M\left(x,V \right)V(t),  \label{eq:1}\\
\dot{x}&=&f\left(x,V\right). \label{eq:2}
\end{eqnarray}
Here, $I$ and $V$ are the current through and the voltage across the device, respectively, $G_M\left( x, V\right)$ stands for the \emph{memory conductance}, memductance for short, $x$ represents  
some internal state variable, and $f\left(x, V\right)$ denotes the \emph{state evolution function}. As demonstrated experimentally~\cite{kim2020experimental}, physical resistive switching memory devices are not (ideal) memristors and therefore their response can be described by the first-order (Eqs.~(\ref{eq:1})-(\ref{eq:2})) or higher-order memristive model~\cite{chua76a}, rather than the memristor model~\cite{chua71a}.

Figure~\ref{fig:1} schematically represents the concept of dynamical attractors~\cite{Pershin_2019}. A necessary requirement for the presence of a dynamical attractor, for a device periodically stimulated by a pulse sequence, is the dependence of the state evolution function $f(x,V)$ in Eq.~(\ref{eq:2}) on $x$~\cite{Pershin_2019}. In first-order memristive systems, the point $x=x_a$ is a fixed-point attractor if Eqs.~(3) and (5) in Ref.~\cite{Pershin_2019} are simultaneously satisfied. An essential characteristic of a fixed-point attractor is the convergence of trajectories towards the attractor from its basin of attraction. During this process, the system loses information about its initial state, demonstrating fading memory~\cite{ascoli2016}. 
Importantly, dynamical attractors are not limited to devices described by first-order models~\cite{Pershin_2019}.

This paper is organized as follows. 
In Sec.~\ref{sec:2}, we present the VTEAM model and derive a time-averaged trajectory solution for the internal state of a resistive switching memory device,
under pulse train excitation. 
Sec.~\ref{sec:3a} outlines the objectives of the state programming operation, while the remainder of Sec.~\ref{sec:3} describes the proposed theoretical methodology for synthesizing the most energy-efficient state programming protocol (Sec.~\ref{sec:3b}) and the most energetically-favorable programming scheme under a limited time budget (Sec.~\ref{sec:3c}).
Joule losses across memristive devices under programming are assessed in Sec.~\ref{sec:4}.
The discussion of our results can be found in Sec.~\ref{sec:5}.  The paper concludes with a conclusion.

\section{Model and trajectory} \label{sec:2}

\subsection{VTEAM model}

Our study is based on the VTEAM  model~\cite{Kvatinsky2013}. In this model, Eq.~(\ref{eq:2}) has the following form:
\begin{numcases}{\frac{\textnormal{d} x}{\textnormal{d}t} =f(x,V)=} 
      k_{off}\left(\frac{V}{v_{off}}-1\right)^{\alpha_{off}} f_{off}(x), & $0<v_{off}<V$, \label{eq:VTEAM_RESET} \\
      0, & $v_{on}<V<v_{off}$, \label{eq:VTEAM_read}\\
      k_{on}\left(\frac{V}{v_{on}}-1\right)^{\alpha_{on}} f_{on}(x), & $V<v_{on}<0$, \label{eq:VTEAM_SET}
\end{numcases}
where $k_{off}>0$, $k_{on}<0$, $\alpha_{off}>0$, $\alpha_{on}>0$ are constant parameters, 
whereas $f_{off}(x)$ and $f_{on}(x)$ are window functions, restricting the allowable variation range of the state variable $x$, \textcolor{black}{that, in some physical devices, represents} \textcolor{black}{the normalized distance between
the tip of a conductive filament and the electrode opposite to the one, the filament
itself originated from}, to the closed interval $[x_{on},x_{off}]$, with $x_{on}=0$ and $x_{off}=1$, and reading as 

\begin{equation}
    f_{off}(x)=1-x, \nonumber
\end{equation}
and
\begin{equation}
    f_{on}(x)=x, \nonumber
\end{equation}
respectively. 
According to the state evolution function $f(x,V)$ (Eqs. (\ref{eq:VTEAM_RESET})-(\ref{eq:VTEAM_SET})), the device undergoes RESET/SET resistive switching transitions \textcolor{black}{ when the voltage V, applied across the device,} is above/below a positive/negative threshold level $v_{off}$/$v_{on}$. 


The memductance $G_M(x,V)$, which appears in the generalized Ohm's law (\ref{eq:1}),  is selected as
\begin{equation}
    G_M(x)=G_{max}+(G_{min}-G_{max})x,
\end{equation}
where $G_{min}$ and $G_{max}$ are the minimum 
and maximum 
conductance 
levels, corresponding to $x_{off}=1$ and $x_{on}=0$, respectively. 

\textcolor{black}{We note that VTEAM~\cite{Kvatinsky2013} is a phenomenological model whose parameters are obtained by fitting measurement results. In particular, Ref.~\cite{Kvatinsky2013} reports parameters for several experimental resistive switching memory devices.}


\subsection{Trajectory}

\textcolor{black}{In this work, we focus on programming devices with square pulses (as illustrated in Fig.~\ref{fig:1}(a)), although other pulse shapes can also be used for this purpose. Square pulses are experimentally simpler to implement, more straightforward to analyze, and, when sufficient programming time is available, they are always optimal for switching~\cite{slipko2025low}.}

In the limit of narrow pulses, the time-averaged dynamics of pulse-driven memristors can be described by the equation~\cite{Slipko19a}
\begin{equation}\label{eq:average}
  \dot{\bar{x}}(t)=\frac{1}{T}\left(f(\bar{x},V_+)\tau_++f(\bar{x},V_-)\tau_-\right),
\end{equation}
where $T$ is the pulse sequence period, $V_\pm$ are the pulse amplitudes, and $\tau_\pm$ are the pulse durations, see Fig.~\ref{fig:1}. Defining
\begin{equation}
A =  k_{off}\left(\frac{V_+}{v_{off}}-1\right)^{\alpha_{off}} \frac{\tau_+}{T} \label{eq:A}
\end{equation}
and
\begin{equation}
B = k_{on}\left(\frac{V_-}{v_{on}}-1\right)^{\alpha_{on}} \frac{\tau_-}{T}\;, \label{eq:B}
\end{equation}
Eq.~(\ref{eq:average}) is rewritten as 
\begin{equation}
    \dot{\bar{x}}(t)=A\left(1-\bar{x}\right)+B\bar{x}. \label{eq:eq}
\end{equation}
The solution of Eq.~(\ref{eq:eq}) is given by
\begin{equation}
    \bar{x}(t)=\frac{A}{A-B}+\left( x_0-\frac{A}{A-B}\right)e^{-(A-B)t}, \label{eq:analytsol}
\end{equation}
where $x(t=0)=x_0$ denotes the initial condition (state). 
In Eq.~(\ref{eq:analytsol}), the first term represents the location of a stable 
fixed point, 
which we denote by
\begin{equation}
    x_a \equiv \frac{A}{A-B} \;, \label{eq:xa}
\end{equation} 
while the second term describes the relaxation process from the initial state $x_0$ towards the equilibrium point $x_a$. 
The corresponding relaxation time is defined via $\tau_r\equiv (A-B)^{-1}$. Note that the memristor state trajectory $x(t)$ is fully and uniquely defined by the constants $A$ and $B$, and by the initial condition $x_0$. Here we emphasize that $B<0$, as it follows from Eqs.~(\ref{eq:B}) and from the negative sign of $k_{on}$ (see the line directly following Eq.~(\ref{eq:VTEAM_SET})).

In principle, the Joule losses across the device are defined by the exact trajectory $x(t)$ of the state variable and the applied time-dependent voltage $V(t)$. 
However, in the limit of narrow pulses, the exact trajectory $x(t)$ may be well  approximated by the time-averaged trajectory $\bar{x}(t)$. 
Under these circumstances the Joule losses can be estimated accurately via 
\begin{equation}
    Q[x(t),V(t)]\approx\frac{1}{T}\left(V_+^2\tau_++V_-^2\tau_- \right)\int\limits_0^{t_f}G_M(\bar{x}(t))\textnormal{d}t \;,
    \label{eq:Q}
\end{equation}
where $t_f$ denotes the programming time. Eq.~(\ref{eq:Q}) forms the basis for the optimization strategy  proposed in this work. 

\section{Optimal trajectory design} \label{sec:3}

\subsection{Objectives of state programming} \label{sec:3a}

As programming objectives, we select the following two parameters: the location of the stable fixed point, $x_a$, and the point-to-point amplitude $\varepsilon$ of long-time oscillations of $x(t)$ about $x_a$. The latter parameter establishes the accuracy of the state programming. In the following, we demonstrate that the parameters $A$ and $B$ in Eqs.~(\ref{eq:A}) and (\ref{eq:B}) are determined by these two programming objectives. For the reader's ease, Fig.~\ref{fig:4}(c) below displays the meaning of the parameters $x_a$ and $\varepsilon$.

Now, consider the amplitude of oscillations of $x(t)$ about  $x_a$ in the long-time limit. Accounting for Eq.~(\ref{eq:2}), which describes the full dynamics, the changes in the internal state due to positive and negative pulses can be approximately expressed as~\cite{pershin_2019_bif_anal_TaO_mod} $\Delta_+= f(x_a,V_+)\tau_+$ and $\Delta_-=f(x_a,V_-)\tau_-$, such that $\Delta_++\Delta_-=0$. The parameter $\epsilon$ is introduced as $\epsilon\equiv |\Delta_{\pm}|$. For the VTEAM model, taking into account Eqs.~(\ref{eq:VTEAM_RESET}), (\ref{eq:VTEAM_SET}), (\ref{eq:A}) and (\ref{eq:B}), one finds
\begin{eqnarray}
    AT(1-x_a)&=&\epsilon \;  ,\\
    -BTx_a&=&\epsilon \; .
\end{eqnarray} 
Consequently, the parameters $A$ and $B$ can be  expressed as
\begin{eqnarray}
    \label{eq:A1} A&=&\frac{\epsilon}{T(1-x_a)} \;  ,\\
    \label{eq:B1} B&=&-\frac{\epsilon}{Tx_a} \; .
\end{eqnarray} 

Once the values of $A$ and $B$ are set, Eqs.~(\ref{eq:A}) and (\ref{eq:B}) provide relationships between the amplitudes and durations of negative and positive pulses. In particular, the durations can be expressed as 
\begin{eqnarray}
 \label{eq:tp}   \tau_+(V_+)&=&\frac{\epsilon}{ k_{off}\left(\frac{V_+}{v_{off}}-1\right)^{\alpha_{off}}(1-x_a)} \;  ,\\ \label{eq:tm}
    \tau_-(V_-)&=&-\frac{\epsilon}{ k_{on}\left(\frac{V_-}{v_{on}}-1\right)^{\alpha_{on}}x_a} \;  .
\end{eqnarray}
Eqs.~(\ref{eq:tp}) and (\ref{eq:tm}) demonstrate that to attain greater precision (i.e., smaller $\epsilon$), one must employ shorter pulses.

\subsection{The most energy efficient programming scheme} \label{sec:3b}

This subsection assumes that the pulse period $T$, another parameter in Eqs.~(\ref{eq:A1}) and (\ref{eq:B1}), is long enough to implement the most energy-efficient sequences. The opposite case is considered in Sec.~\ref{sec:3c} below.

With $A$ and $B$ determined (refer to Eqs.~(\ref{eq:A1}) and (\ref{eq:B1})), the pre-integral factor in Eq.~(\ref{eq:Q}) remains the sole element available for minimization. Our energy loss minimization procedure applies separately to pulses of positive and negative polarity. 
By substituting $\tau_+$ from Eq.~(\ref{eq:tp}) into Eq.~(\ref{eq:Q}), for the positive pulse, the optimization problem reduces to
\begin{equation}
    \hat{V}_+=\underset{v_{off}\leq V_+\leq V_+^{max}}{\textnormal{Argmin}}\;\frac{V_+^2}{\left(\frac{V_+}{v_{off}}-1\right)^{\alpha_{off}}}\; , \label{eq:Vp}
\end{equation}
where $V_+^{max}$ is the most positive voltage to which the device may be subjected, and the hat symbol denotes the \textit{optimal} pulse amplitude.
The minimization of Joule losses under RESET switching transitions results in the requirement to choose the height of the negative pulses according to

\begin{equation}
    \hat{V}_-=\underset{V_-^{max}\leq V_-\leq v_{on}}{\textnormal{Argmin}}\;\frac{V_-^2}{\left(\frac{V_-}{v_{on}}-1\right)^{\alpha_{on}}}\; , \label{eq:Vn}
\end{equation}
where $V_-^{max}$ is the most negative voltage applicable across the device. \textcolor{black}{$V_{+/-}^{max}$ should be selected taking into account the largest voltage or voltages available in the circuit and the operational voltage stress thresholds for the
device.}

\begin{figure*}[t]
    \centering
    \includegraphics[width=0.3\linewidth]{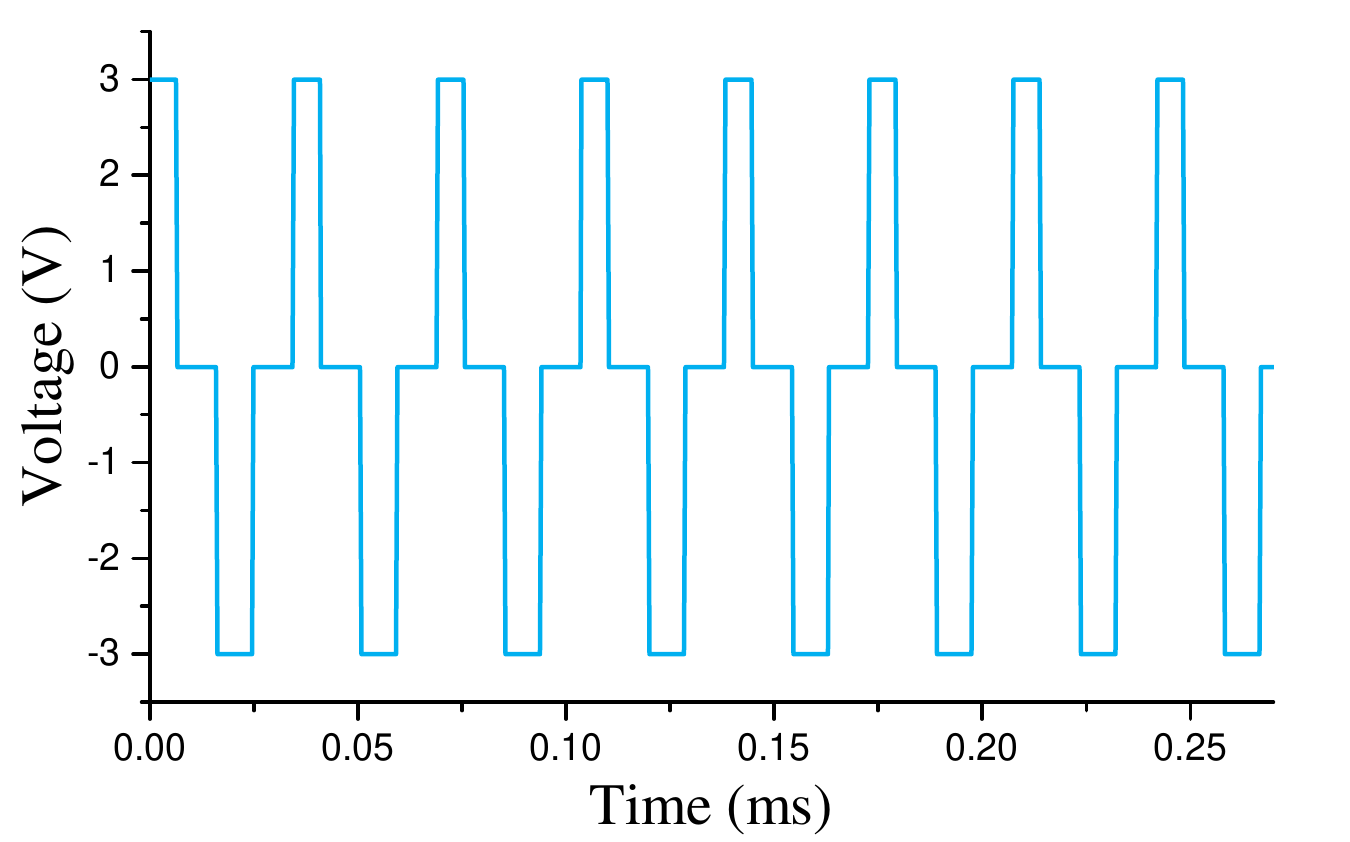}
    \includegraphics[width=0.3\linewidth]{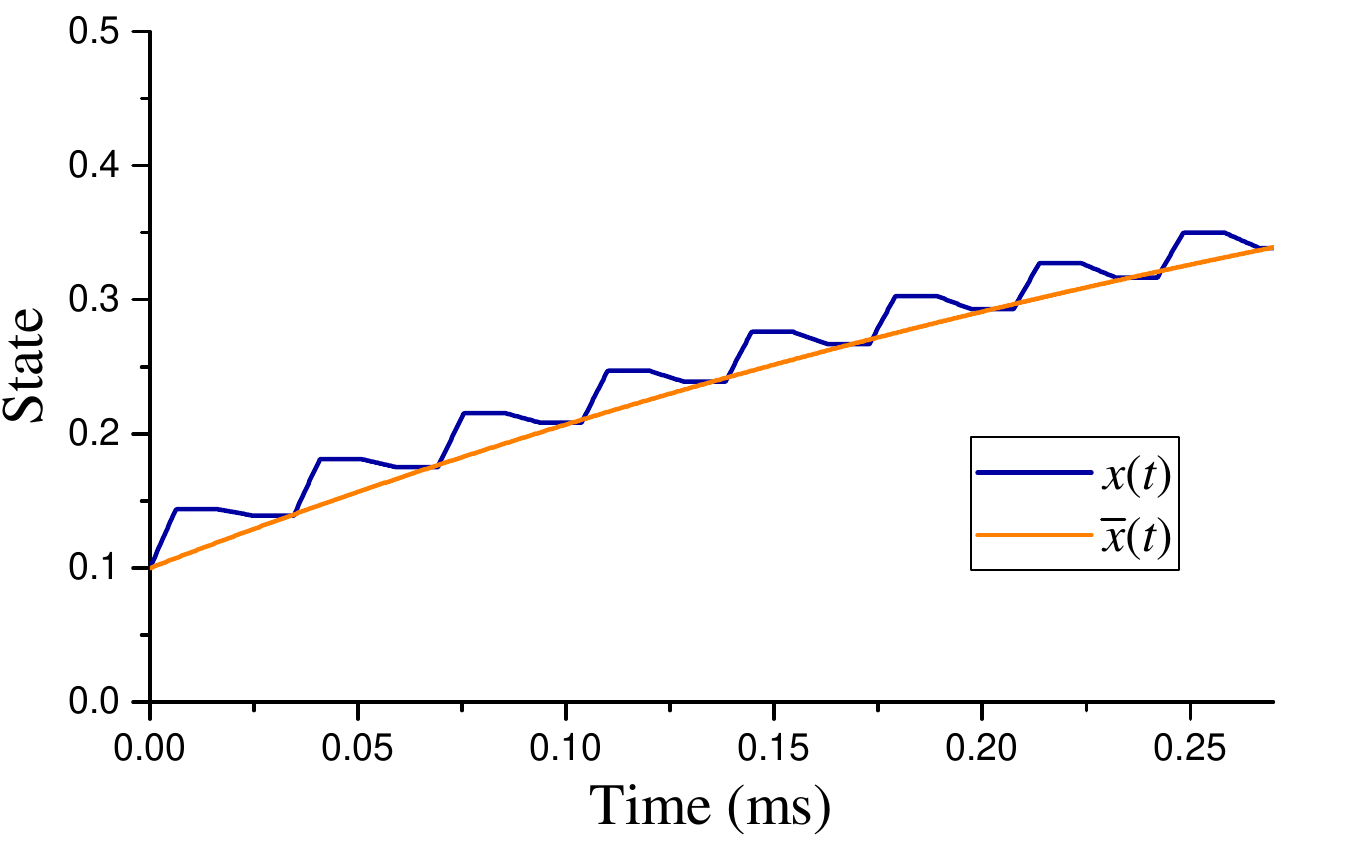}
    \includegraphics[width=0.3\linewidth]{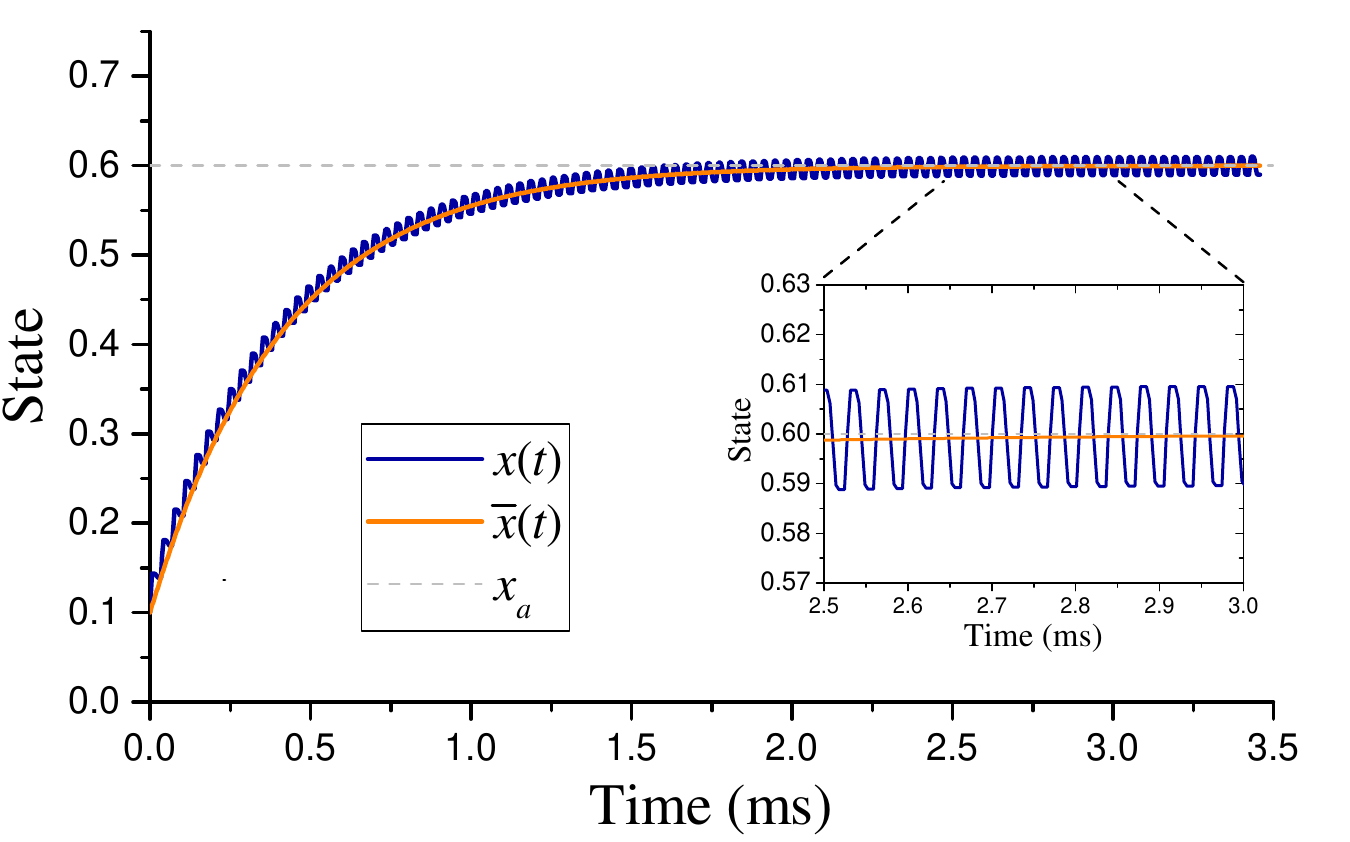}\\
    (a)\hspace{5cm}(b)\hspace{5cm}(c)
    \caption{Example of optimal control based upon the following parameter setting: $v_{off/on}=\pm 1$~V,  $\alpha_{off}=3$, $\alpha_{on}=2$, $k_{off/on}=\pm 10^3$, $V_{\pm}^{max}=\pm 3$~V, $x_0=0.1$, $x_a=0.6$, $\epsilon=0.02$, $\tau_0=10$~{\textmu}s. 
    Here, plot (a) shows the control voltage pulse sequence, where $\hat{V}_{+}=V_+^{max}=3$V and $\hat{V}_{-}=V_-^{max}=-3$V according to Eqs. (\ref{eq:amplitudes+}) and (\ref{eq:amplitudes-}), respectively, 
    while 
    $\tau_{+}=6.3$~{\textmu}s and $\tau_{-}=8.3$~{\textmu}s as follows in turn from Eqs. (\ref{eq:tp}) and (\ref{eq:tm}). The pulse sequence admitted then an input period $T$ of $34.6$~{\textmu}s, 
    as computed via Eq. (\ref{eq:24}). 
    Plots (b) and (c) depict the memristor state response to the pulse train in (a) over the initial phase and across the entire programming time, respectively. In (b) and (c), the smooth line represents the analytical solution according to Eq.~(\ref{eq:analytsol}).} %
    \label{fig:3}
\end{figure*}

\begin{figure*}[t]
    \centering
    \includegraphics[width=0.3\linewidth]{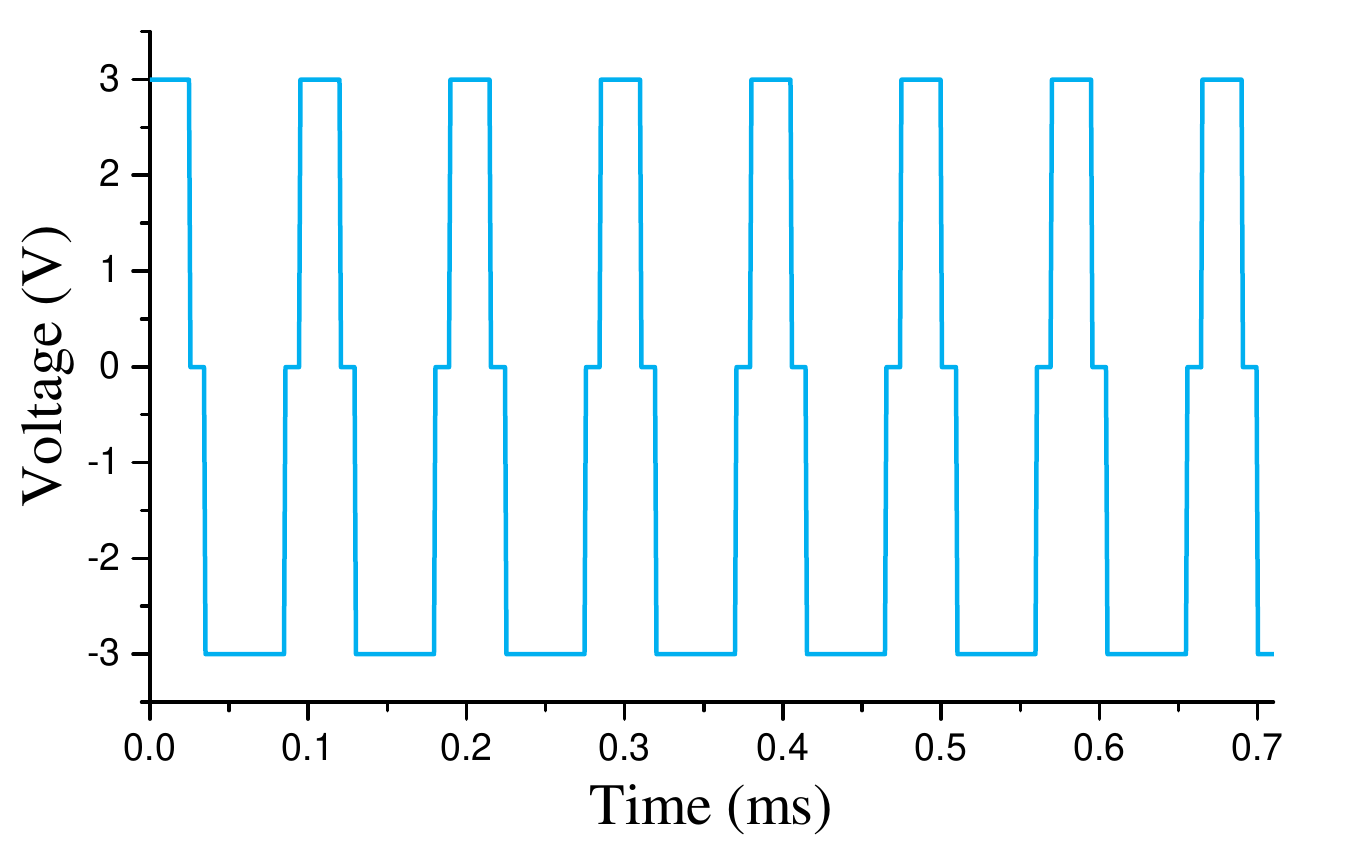}
    \includegraphics[width=0.3\linewidth]{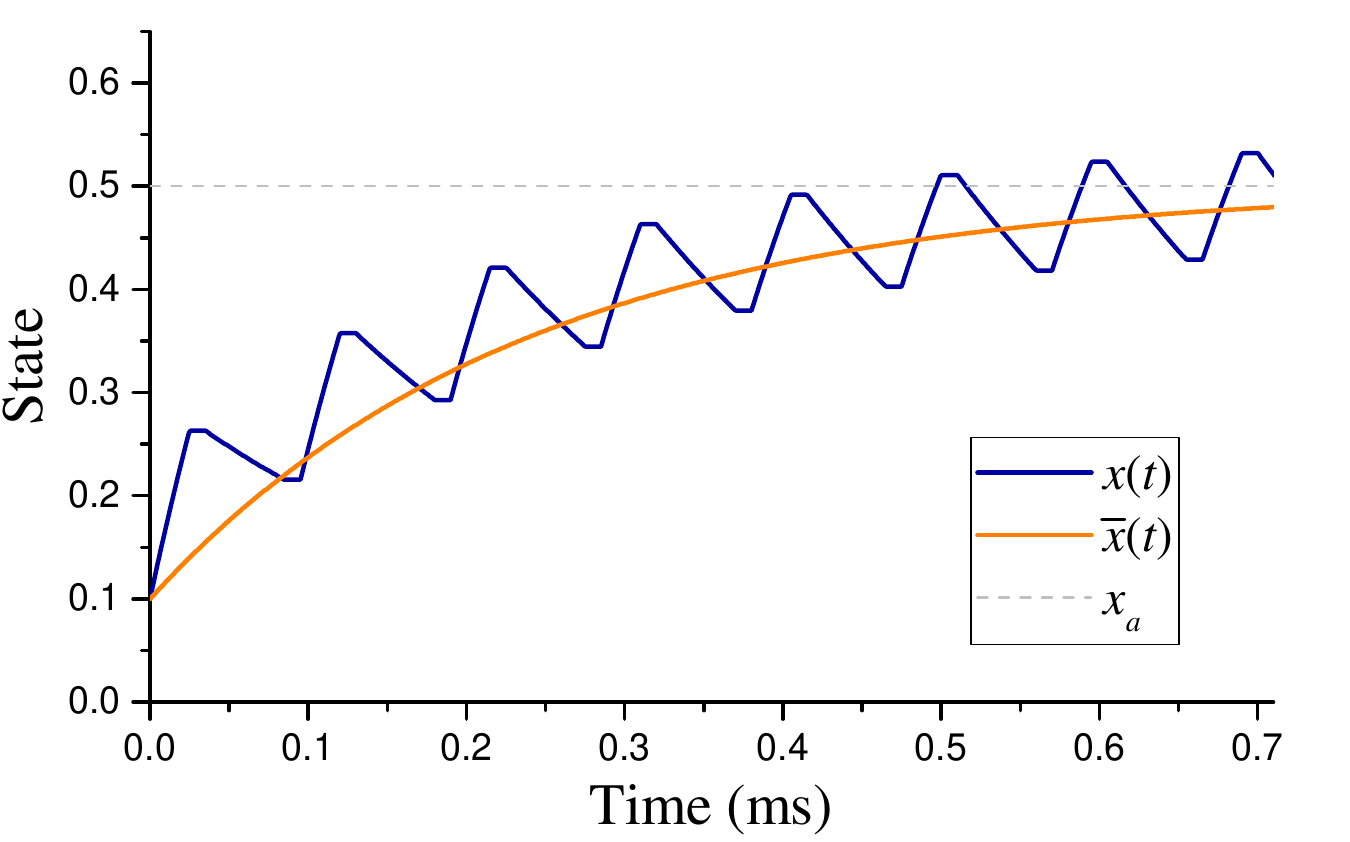}
    \includegraphics[width=0.3\linewidth]{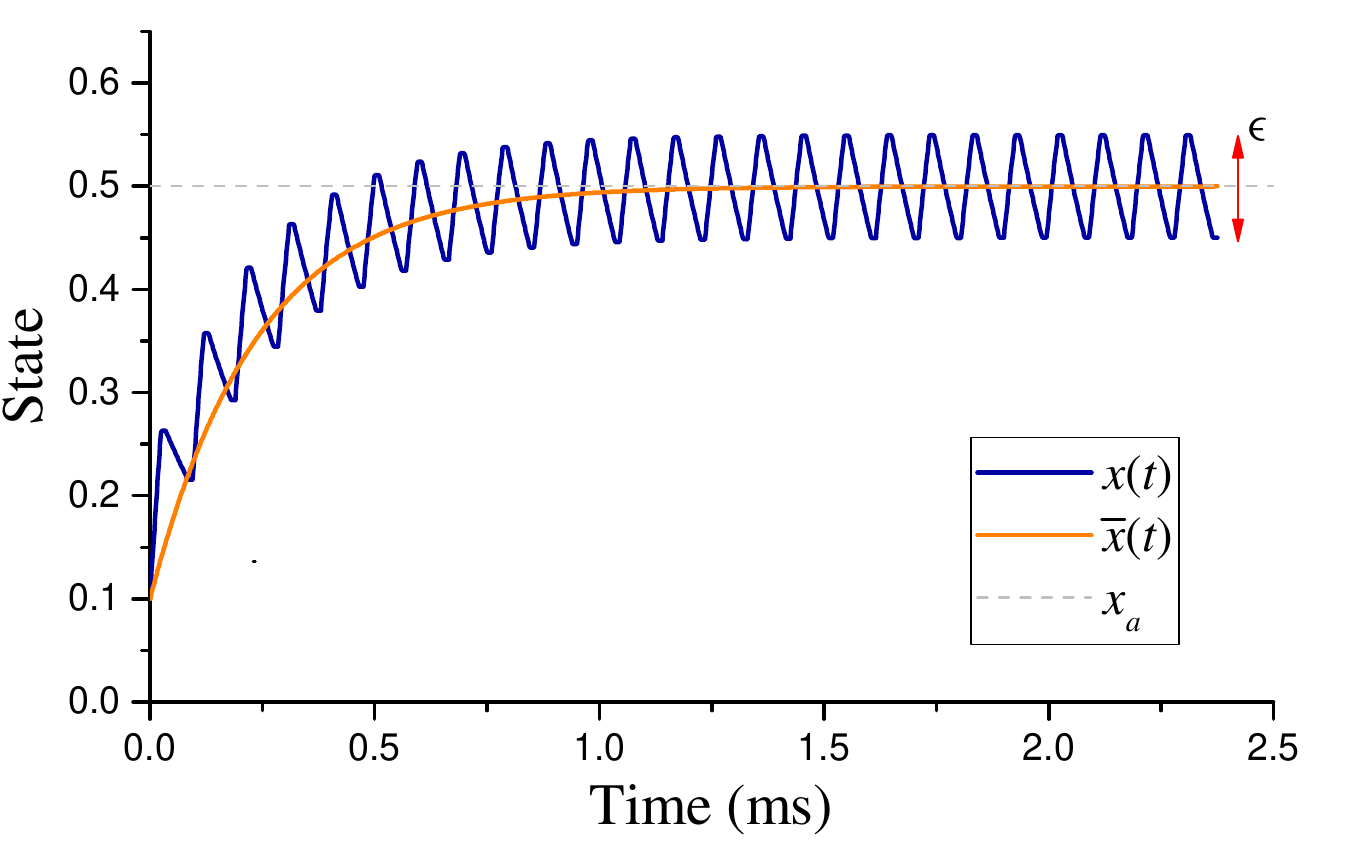}\\
    (a)\hspace{5cm}(b)\hspace{5cm}(c)
    \caption{Example of optimal control based upon the same parameter setting as in Fig.~\ref{fig:3}, except for $x_a$, and $\epsilon$, here taken equal to $0.5$ and $0.1$, respectively. 
    See the caption of Fig. \ref{fig:3} for the significance of the curves in plots (a), (b), and (c). 
    For this simulation, according to Eqs.~(\ref{eq:amplitudes+}) and (\ref{eq:amplitudes-}) $\hat{V}_{+}$ and $\hat{V}_{-}$ were respectively set to 
    $V_+^{max}=3$V and $V_-^{max}=-3$V, as in Fig.~\ref{fig:3}. 
    %
    As a result, using the formulas 
    (\ref{eq:tp}) and (\ref{eq:tm}), $\tau_{+}$ and $\tau_{-}$ were in turn chosen equal to $25$~{\textmu}s and $50$~{\textmu}s. From equation (\ref{eq:24}), the pulse sequence period $T$ was thus found to amount to 95~{\textmu}s.}
    \label{fig:4}
\end{figure*}

\begin{figure*}[ht]
    \centering
    \includegraphics[width=0.3\linewidth]{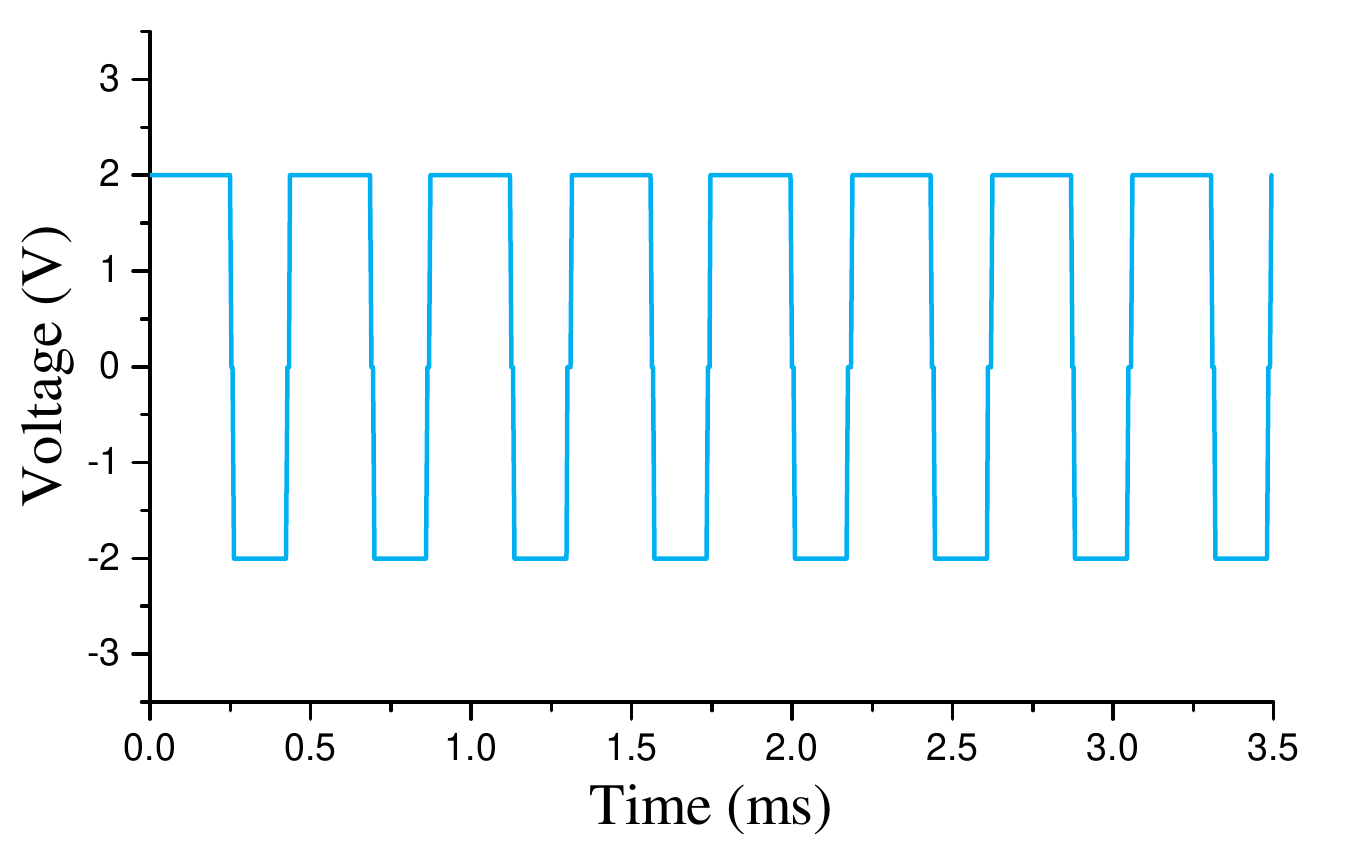}
    \includegraphics[width=0.3\linewidth]{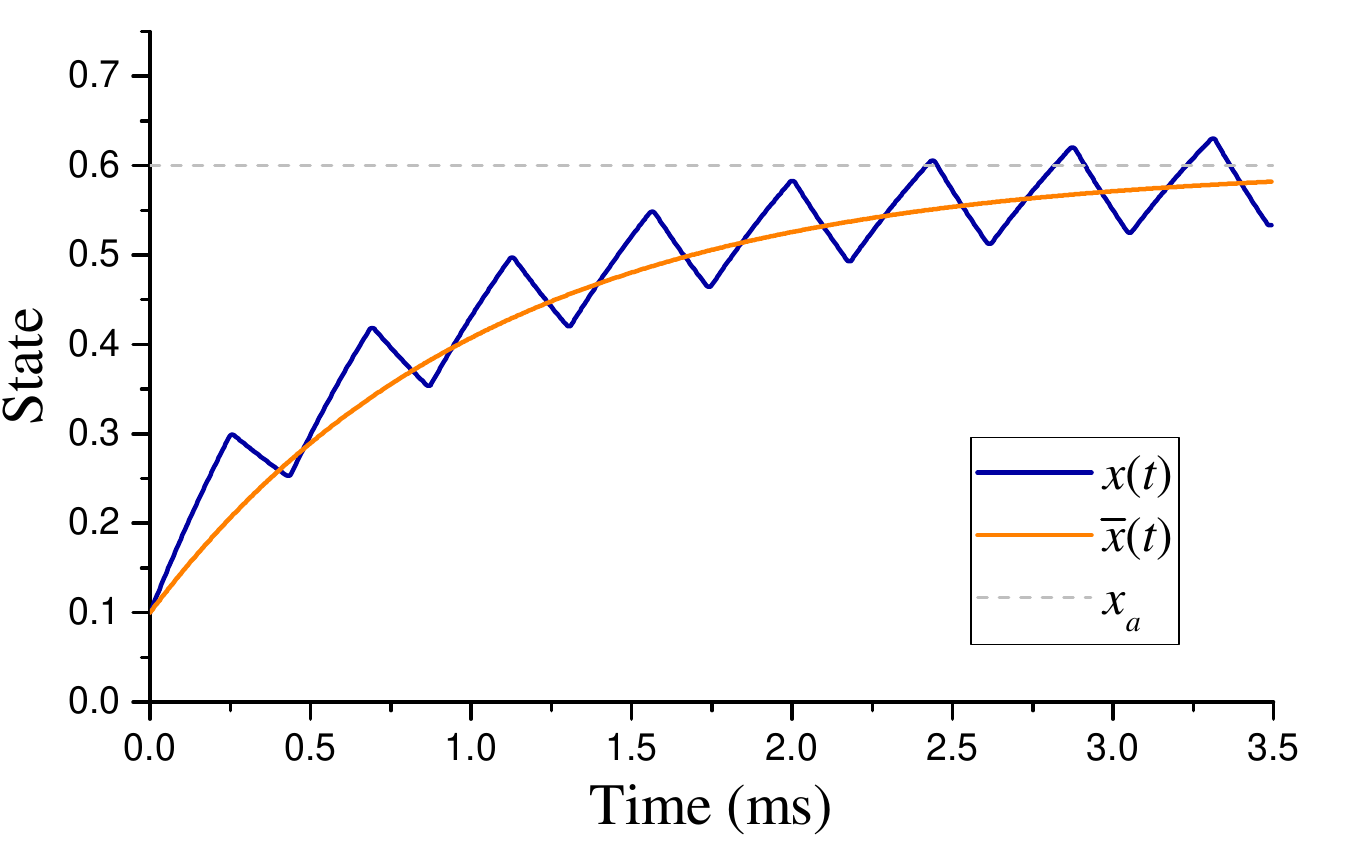}
    \includegraphics[width=0.3\linewidth]{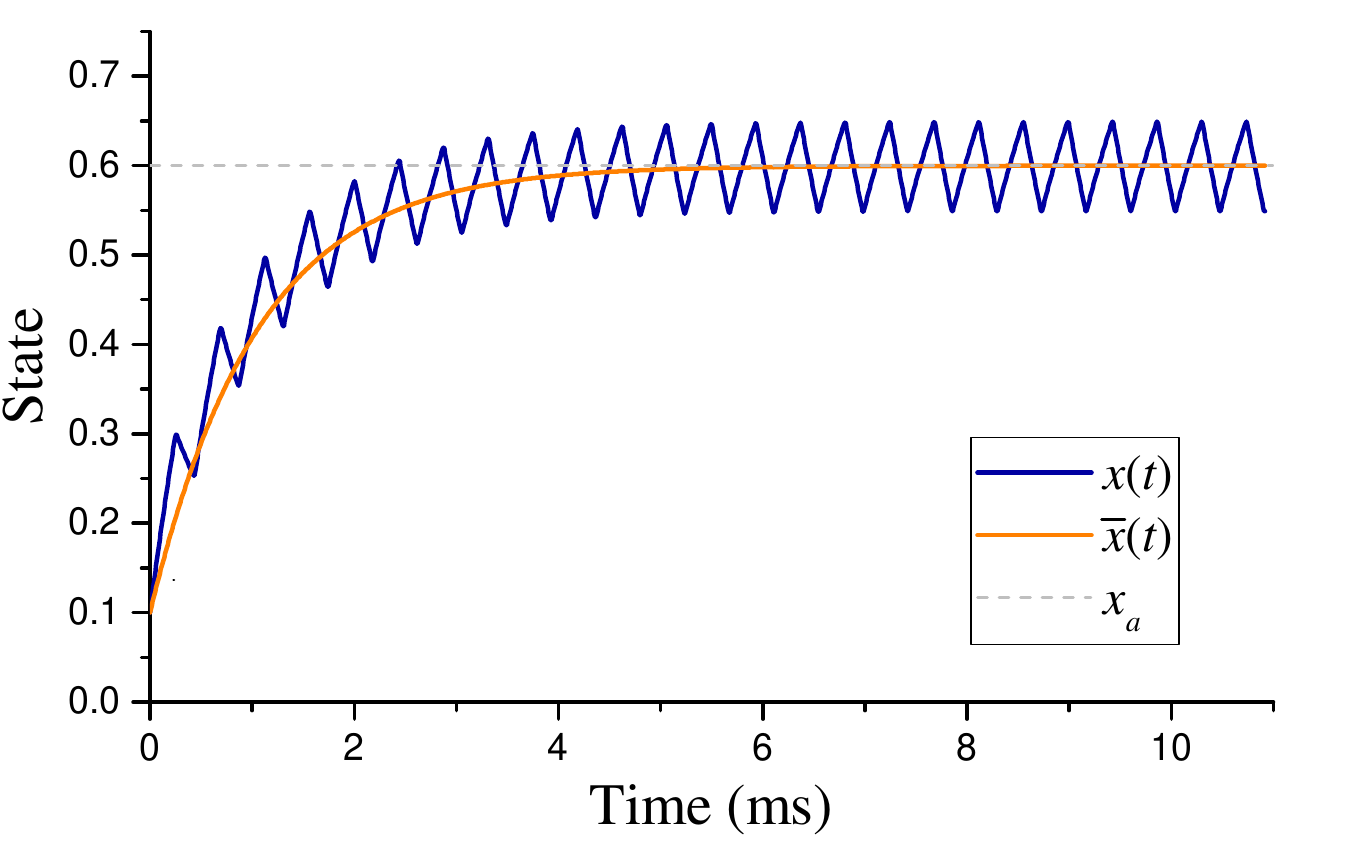}\\
    (a)\hspace{5cm}(b)\hspace{5cm}(c)
    \caption{Example of optimal control based upon the same parameter setting as in Fig.~\ref{fig:3}, except for $\alpha_{off/on}$, and $\epsilon$, here set equal to $1$ and $0.1$, respectively.  
    Here the inequalities $V_{+}^{max}>2\,v_{off}/(2-\alpha_{off})$ and $V_{-}^{max}<2v_{on}/(2-\alpha_{on})$ hold true. 
    Refer to the caption of Fig. \ref{fig:3} for information on the traces in plots (a), (b), and (c).  
    For this simulation Eqs. (\ref{eq:amplitudes+}) and (\ref{eq:amplitudes-}) respectively set the optimal RESET and SET pulse amplitudes, i.e., in turn, $\hat{V}_{+}$ and $\hat{V}_{-}$, to
    $2\,v_{off}/(2-\alpha_{off})=2$V and $2\,v_{on}/(2-\alpha_{on})=-2$V, respectively. 
    As a result,  $\tau_{+}$ and $\tau_{-}$ were set to $250$~{\textmu}s and $167$~{\textmu}s, on the basis of Eqs. (\ref{eq:tp}) and (\ref{eq:tm}), respectively. Therefore, $T$ was found to be equal to $437$~{\textmu}s 
    according to equation (\ref{eq:24}).}
    \label{fig:5}
\end{figure*}

Eq.~(\ref{eq:Vp}) leads to
\begin{equation}
\hat{V}_{+} = 
\begin{cases}
\min\left\{ \dfrac{2}{2 - \alpha_{\text{off}}} v_{\text{off}},\ V_{+}^{\text{max}} \right\}, & \text{for } \alpha_{\text{off}} < 2, \\
V_{+}^{\text{max}}, & \text{for } \alpha_{\text{off}} \geq 2.
\end{cases} \label{eq:amplitudes+}
\end{equation}
 Similarly,
\begin{equation}
\hat{V}_{-} = 
\begin{cases}
\max\left\{ \dfrac{2}{2 - \alpha_{\text{on}}} v_{\text{on}},\ V_{-}^{\text{max}} \right\}, & \text{for } \alpha_{\text{on}} < 2, \\
V_{-}^{\text{max}}, & \text{for } \alpha_{\text{on}} \geq 2.
\end{cases} \label{eq:amplitudes-}
\end{equation}

Eqs.~(\ref{eq:tp}) and (\ref{eq:tm}) enable the determination of pulse widths for given pulse amplitudes (refer to Eqs.~(\ref{eq:amplitudes+}) and (\ref{eq:amplitudes-})). The pulse period $T$ can be found as follows:
\begin{equation}
    T_{opt}=\tau_+(\hat{V}_+)+\tau_-(\hat{V}_-)+2\tau_0,
    \label{eq:24}
\end{equation}
where $2\tau_0$ is the total duration of the time interval across which $V=0$ within the pulse period. Our findings are applicable to both  symmetric (as shown in Fig.~\ref{fig:1}(a)) and asymmetric arrangements of alternating pulses within a period; what truly matters is solely the total length of the zero-input time interval, namely, $2\tau_0$. 
Note that $\tau_+(\hat{V}_+)+\tau_-(\hat{V}_-)$ is the shortest period of the optimal pulse sequence.

Figs. \ref{fig:3}-\ref{fig:5} illustrate examples of optimal sequences obtained through the method outlined in the current subsection.

\begin{figure}
    \centering
    (a) \includegraphics[width=0.45\textwidth]{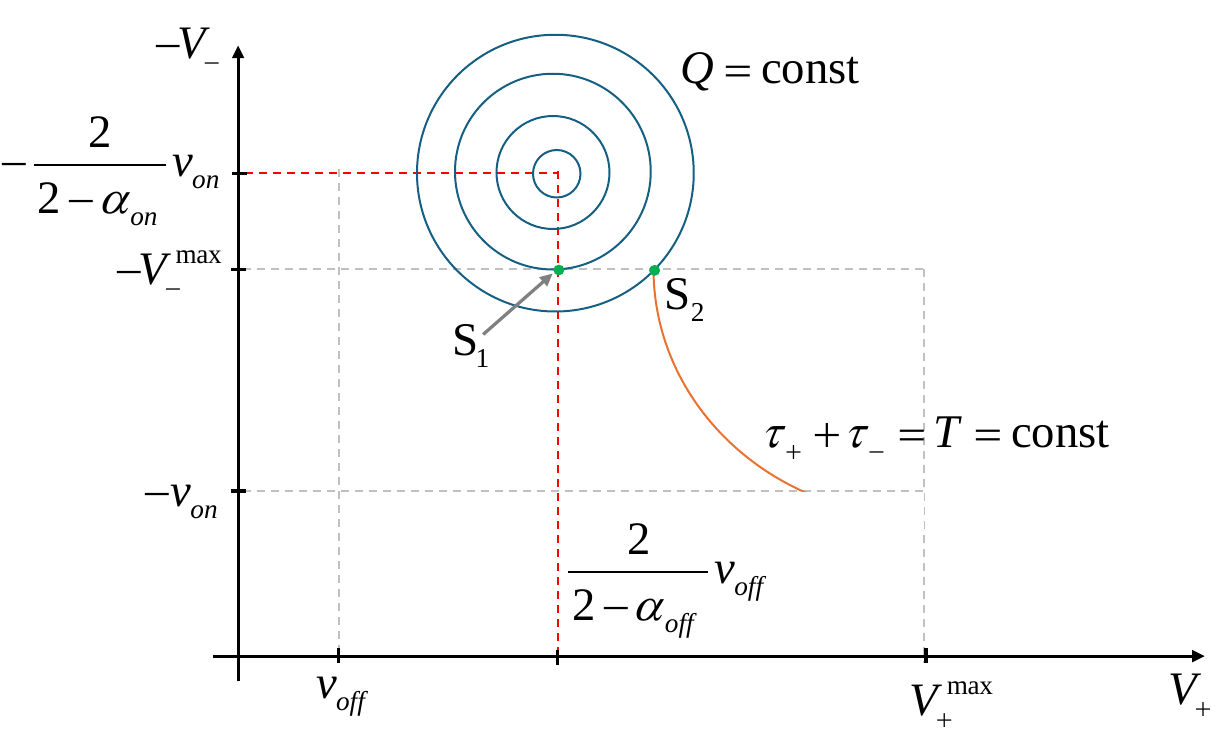} \hspace{0.2cm}
    (b) \includegraphics[width=0.45\textwidth]{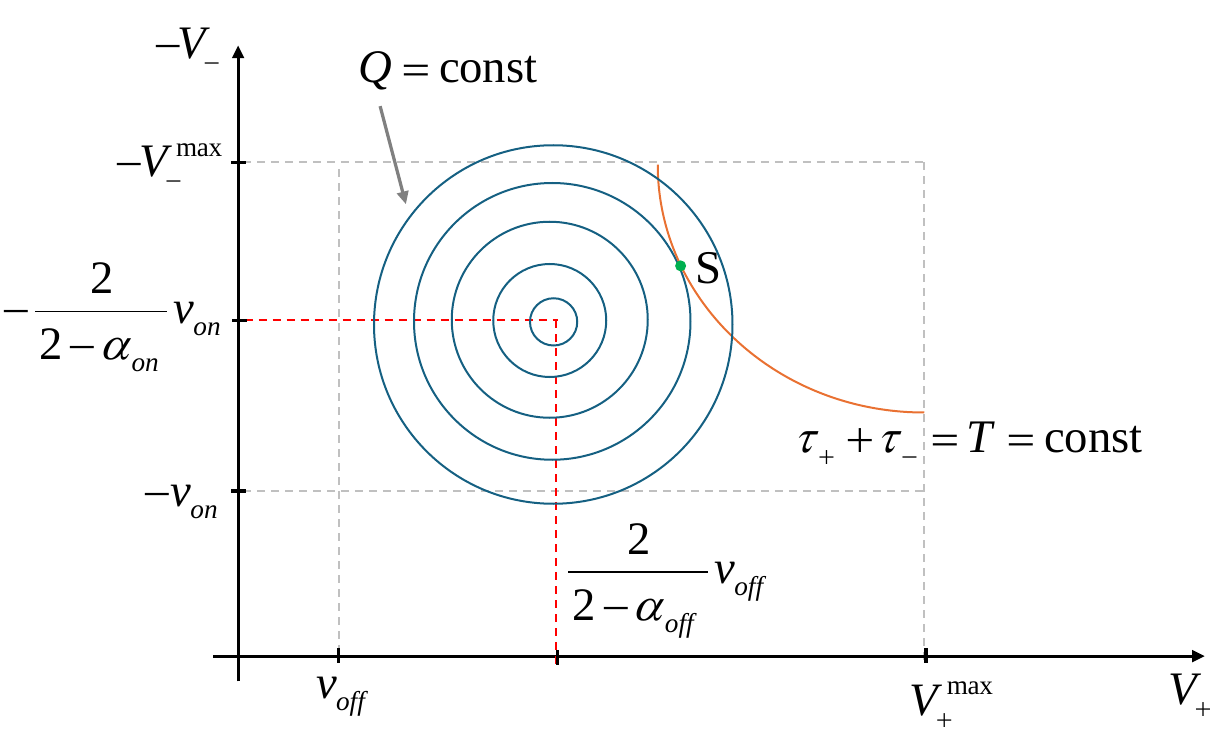}
    \caption{Minimizing Joule losses in the presence of constraints on pulse amplitudes and widths in the case where $\alpha_{\text{off}} < 2$ and $\alpha_{\text{on}} < 2$. The allowable variation ranges for $V_+$ and for $-V_-$ are $[v_{off},V_{+}^{max}]$ and $[-v_{on},-V_{-}^{max}]$, respectively. (a) A schematic illustration of the scenario where $V_+^{max}>2v_{off}/(2-\alpha_{off})$ and $|V_-^{max}|<2|v_{on}|/(2-\alpha_{on})$. (b) A schematic illustration of the scenario where  $V_+^{max}>2\,v_{off}/(2-\alpha_{off})$ and $|V_-^{max}|>2|v_{on}|/(2-\alpha_{on})$. 
    It is worth observing that, when both inequalities $\alpha_{\text{on}}<2$ and $\alpha_{\text{off}}<2$ apply, the contours of the function $Q=Q(V_+,V_-)$ are closed loci, which originates from the fact that in situations of this kind there always exist  finite positive and negative pulse amplitudes, in turn $V_+$ and $V_-$, 
    at which Joule losses are minimized under RESET and SET transitions, respectively, in the unconstrained case (refer to Fig. \ref{fig:6}(a)). 
    } 
    \label{fig:2}
\end{figure}

\subsection{The most energy efficient programming scheme under a limited time budget} \label{sec:3c}

In some cases, it may be necessary to restrict the programming time, even if this leads to Joule losses that are higher than the minimal level obtainable with the sequences outlined earlier in subsection~\ref{sec:3b}.

In all possible scenarios, the shortest pulse period is 
\begin{equation}
    T_{min}=\tau_+(V_+^{max})+\tau_-(V_-^{max})\leq T_{opt}
\end{equation}
as the highest pulse amplitude corresponds to the shortest pulse duration (see Eqs.~(\ref{eq:tp}) and (\ref{eq:tm})).
An interesting case is when the desired pulse sequence period $T$ satisfies 
\begin{equation}
     \left. T_{min}<T<T_{opt}\right|_{\tau_0=0}.
     \label{eq:intrestingT}
\end{equation}
In fact, there are two related cases:

\subsubsection{Case 1}
Consider the scenario, where Eqs.~(\ref{eq:amplitudes+}) and (\ref{eq:amplitudes-}) suggest selecting the pulse amplitude pair  
$\{ \hat{V}_+=V_+^{max},\hat{V}_-=2\,v_{\text{on}}/(2 - \alpha_{\text{on}})\}$. 
This scenario occurs if a) $\alpha_{off}\geq2$ or $\alpha_{off}<2$ and $V_{+}^{max}<2v_{off}/(2-\alpha_{off})$ and b) $\alpha_{on}<2$ and $|V_{-}^{max}|>2|v_{on}|/(2-\alpha_{on})$.   
Let us however also assume that for the application of interest, employing the optimal pulse sequence to program the device would exceed the maximum allowable time. This would call for the adoption of a faster programming protocol, so that the pulse train period $T$ could satisfy the constraint Eq.~(\ref{eq:intrestingT}). In this case, the amplitude for the positive pulse is selected as $\hat{V}_+$, $\tau_+$ is computed via (\ref{eq:tp}), and, concurrently, the width of the negative pulse is calculated as the remainder of the pulse sequence period, i.e via 
\begin{equation}
   \tau_-=T-\tau_+(\hat{V}_+). \label{eq:taum2}
\end{equation}
Then, the amplitude of the negative pulse can be found by solving Eq.~(\ref{eq:tm}) for $V_-$, while using $\tau_-$ from Eq.~(\ref{eq:taum2}). 

Let us now consider the scenario, where Eqs.~(\ref{eq:amplitudes+}) and (\ref{eq:amplitudes-}) result in the optimal pulse amplitude pair choice  
%
$\{ \hat{V}_+=2\,v_{\text{off}}/(2 - \alpha_{\text{off}}),\hat{V}_-=V_-^{max}\}$. 
This scenario occurs if a) $\alpha_{on}\geq2$ or $\alpha_{on}<2$ and $|V_{-}^{max}|<2|v_{on}|/(2-\alpha_{on})$ and b) $\alpha_{off}<2$ and $V_{+}^{max}>2\,v_{off}/(2-\alpha_{off})$. 
Now, let us suppose the memristor-centered application requires the implementation of a fast resistive switching scheme, as dictated by the constraint (\ref{eq:intrestingT}), so as to prevent the programming time to exceed a predefined upper bound.  
In a situation of this kind, the amplitude for the negative pulse is selected as $\hat{V}_-$, $\tau_-$ is computed via (\ref{eq:tm}), and, concurrently, the width of the positive pulse is calculated as the remainder of the pulse sequence period, i.e via 
\begin{equation}
   \tau_+=T-\tau_-(\hat{V}_-). \label{eq:taup2}
\end{equation}
Then, the amplitude of the positive pulse can be found by solving Eq.~(\ref{eq:tp}) for $V_+$, while using $\tau_+$ from Eq.~(\ref{eq:taup2}). 
In fact, a schematic diagram of this case is presented in Fig.~\ref{fig:2}(a) when $\alpha_{on}<2$ and $\alpha_{off}<2$. 
Here, the coordinates of the point ``S$_1$'' represent the optimal pulse amplitudes when $\left.\tau_++\tau_-= T_{opt}\right|_{\tau_0=0}$. However, under the constraint, defined through Eq.~(\ref{eq:intrestingT}), the most energetically-favorable 
solution turns out to be inferable from the coordinates of the point  ``S$_2$'', where a constant-value contour of the function $Q(V_+,V_-)=\textnormal{const}$ crosses 
the curve associated with the condition 

\begin{equation}
    \tau_++\tau_-=T 
    \label{eq:constraint}
\end{equation}
along the line $V_-=V_-^{max}$. 
The resulting pulse sequence would induce a larger energy dissipation 
across the device than the optimal one presented in Sec.~(\ref{sec:3b}). This is the price to pay to program the device resistance at a higher rate. 


\subsubsection{Case 2}
Another possibility emerges when Eqs.~(\ref{eq:amplitudes+}) and (\ref{eq:amplitudes-}) suggest setting the amplitudes of the pulse sequence as 
$\{ \hat{V}_+=2\,v_{off}/(2 - \alpha_{\text{off}}), 
\hat{V}_-=2\,v_{on}/(2 - \alpha_{\text{on}})\}$,
which is the case when the inequalities $\alpha_{on}<2$ and $\alpha_{off}<2$ hold true, while, concurrently, $V_{+}^{max}>2\,v_{off}/(2-\alpha_{off})$ and $|V_{-}^{max}|>2\,|v_{on}|/(2-\alpha_{on})$.  
Yet the period $T_{opt}$ of the most energetically-favourable pulse train would be too long to keep the duration of the programming phase below a given application-dependent upper bound.  
The optimization task would then involve the minimization of $Q$, as given in Eq.~(\ref{eq:Q}), under the condition (\ref{eq:constraint}), which may be achieved by recurring to the standard method of the Lagrange multipliers~\cite{arfken2011mathematical}.

Fig.~\ref{fig:2}(b) illustrates the method graphically. The most energetically-favourable solution in the fast programming scheme corresponds to the point ``S'', where a contour of the function $Q(V_+,-V_-)$ is tangent to the curve illustrating the constraint expressed by Eq.~(\ref{eq:constraint})). 
Mathematically, this condition boils down to 
\begin{equation}
    \vec{\nabla}  Q(V_+,V_-)=\lambda \vec{\nabla} \left( \tau_++\tau_- \right), \label{eq:L}
\end{equation}
where $\vec{\nabla}$ represents the gradient operator, and $\lambda$ is the Lagrange multiplier. Eq.~(\ref{eq:L}) may be expanded as 
\begin{equation}
    \frac{\frac{\partial Q(V_+,V_-)}{\partial V_+}}{\frac{\partial \tau_+(V_+)}{\partial V_+}}=
    \frac{\frac{\partial Q(V_+,V_-)}{\partial V_-}}{\frac{\partial \tau_-(V_-) }{\partial V_-}} \; .\label{eq:L1}
\end{equation}
The location of the point ``S'' in Fig.~\ref{fig:2}(b) can be found by solving simultaneously Eqs.~(\ref{eq:constraint}) and (\ref{eq:L1}) for $V_+$ and $V_-$. 
The widths $\tau_+$ and $\tau_-$ of the positive and negative pulses are then computed by means of Eqs. (\ref{eq:tp}) and (\ref{eq:tm}), respectively. 

\section{Joule losses} \label{sec:4}

Having established the optimal programming protocols in Sec.~\ref{sec:3}, we are now in a perfect position for evaluating the Joule losses in the memristive device during the programming phase. Our starting point is Eq.~(\ref{eq:Q}). The integral in Eq.~(\ref{eq:Q}) can be calculated analytically, leading to the closed-form expression
\begin{equation}
I\equiv \int\limits_0^{t_f}G_M(\bar{x}(t))\textnormal{d}t=G_M(x_a)t_f+\left(x_0-x_a\right)\left(G_{min}-G_{max}\right)\left(1-e^{-\frac{t_f}{\tau_r}}\right)\tau_r \;. \;\; \label{eq:Qintegral}
\end{equation}
On the right-hand side of Eq.~(\ref{eq:Qintegral}), the first term is proportional to the programming time $t_f$ and accounts for Joule losses associated with the attractor state. Meanwhile, the second term 
is a finite contribution related to the transient relaxation dynamics occurring over the time scale $\tau_r$. This contribution 
may be either positive or negative, depending on the location of $x_0$ with respect to $x_a$.

Within Eq.~(\ref{eq:Q}), the 
programming time $t_f$ is yet to be determined. Taking into account the exponential behavior given by Eq.~(\ref{eq:analytsol}), the relaxation time $\tau_r=(A-B)^{-1}$ emerges as a suitable time scale to define $t_f$. 
Practically, one should allow for a time span of several relaxation times to ensure the device to attain a location adequately close to the attractor. 
As a result, we suggest choosing $t_f$ according to the specification 
\begin{equation}
    t_f=p \tau_r \;, 
\end{equation} 
where $p$ is a real number generally falling within the range 
$[2,5]$ in most applications.

Here we point out that Eq.~(\ref{eq:Qintegral}) remains 
invariant across various programming protocols (assuming the same trajectory). Variations arise from the pre-integral coefficient in Eq.~(\ref{eq:Q}), which must be addressed individually for different \textcolor{black}{pulse sequences.} In all cases, however, the Joule losses in the memristive device may be accurately evaluated inspecting the closed-form expression 

\begin{equation}
    Q=\left[\frac{\epsilon V^2_+}{ k_{off}\left(\frac{V_+}{v_{off}}-1\right)^{\alpha_{off}}(1-x_a)} -  
    \frac{\epsilon V^2_-}{ k_{on}\left(\frac{V_-}{v_{on}}-1\right)^{\alpha_{on}}x_a} \right] \frac{I}{T}\;,
    \label{eq:Qfinal}
\end{equation}
where $I$ is given by Eq.~(\ref{eq:Qintegral}). Below are a few examples for illustration.

\begin{figure}[t]
    \centering
    (a)
    \includegraphics[width=0.35\columnwidth]{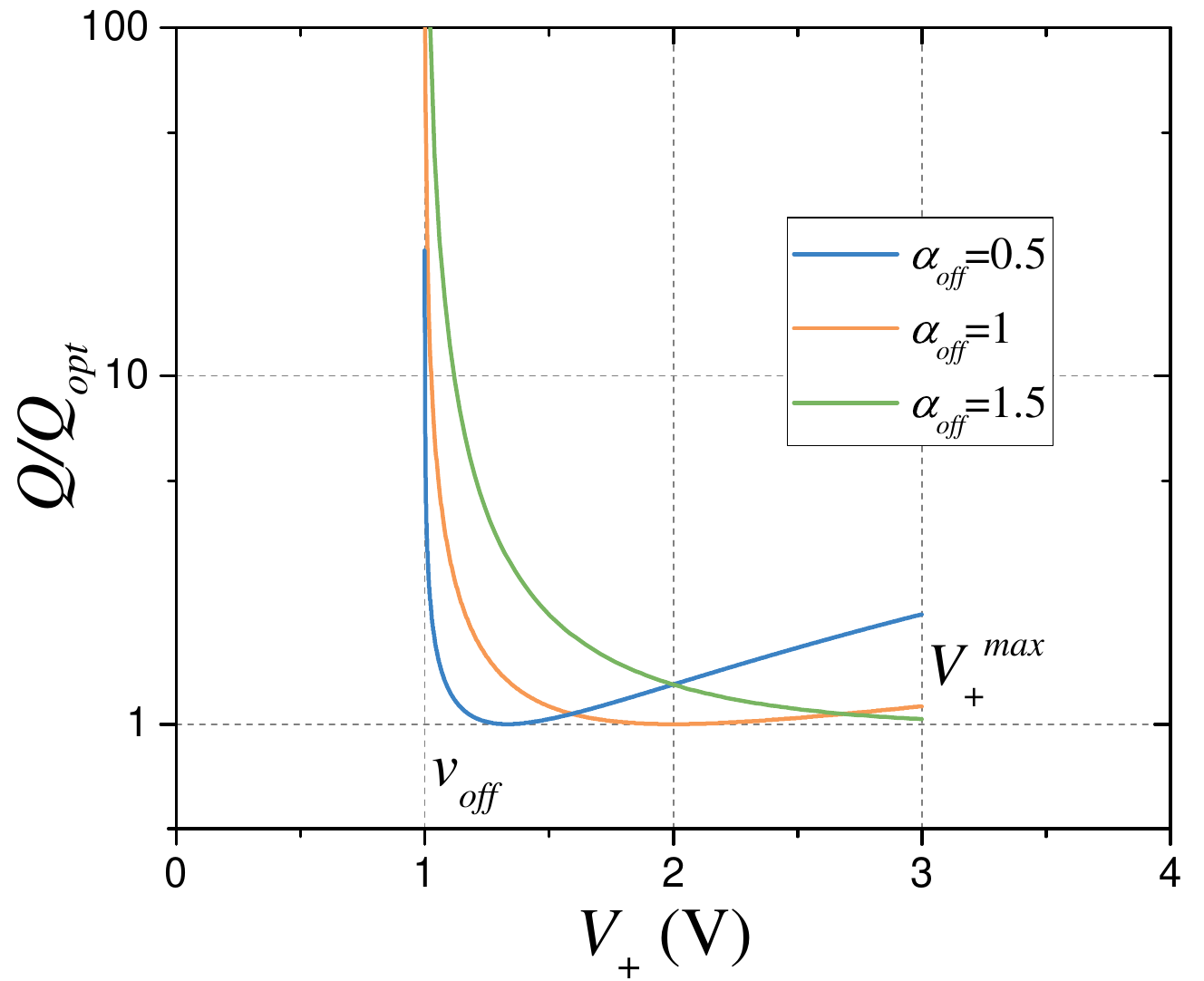}    
    (b)
    \includegraphics[width=0.35\linewidth]{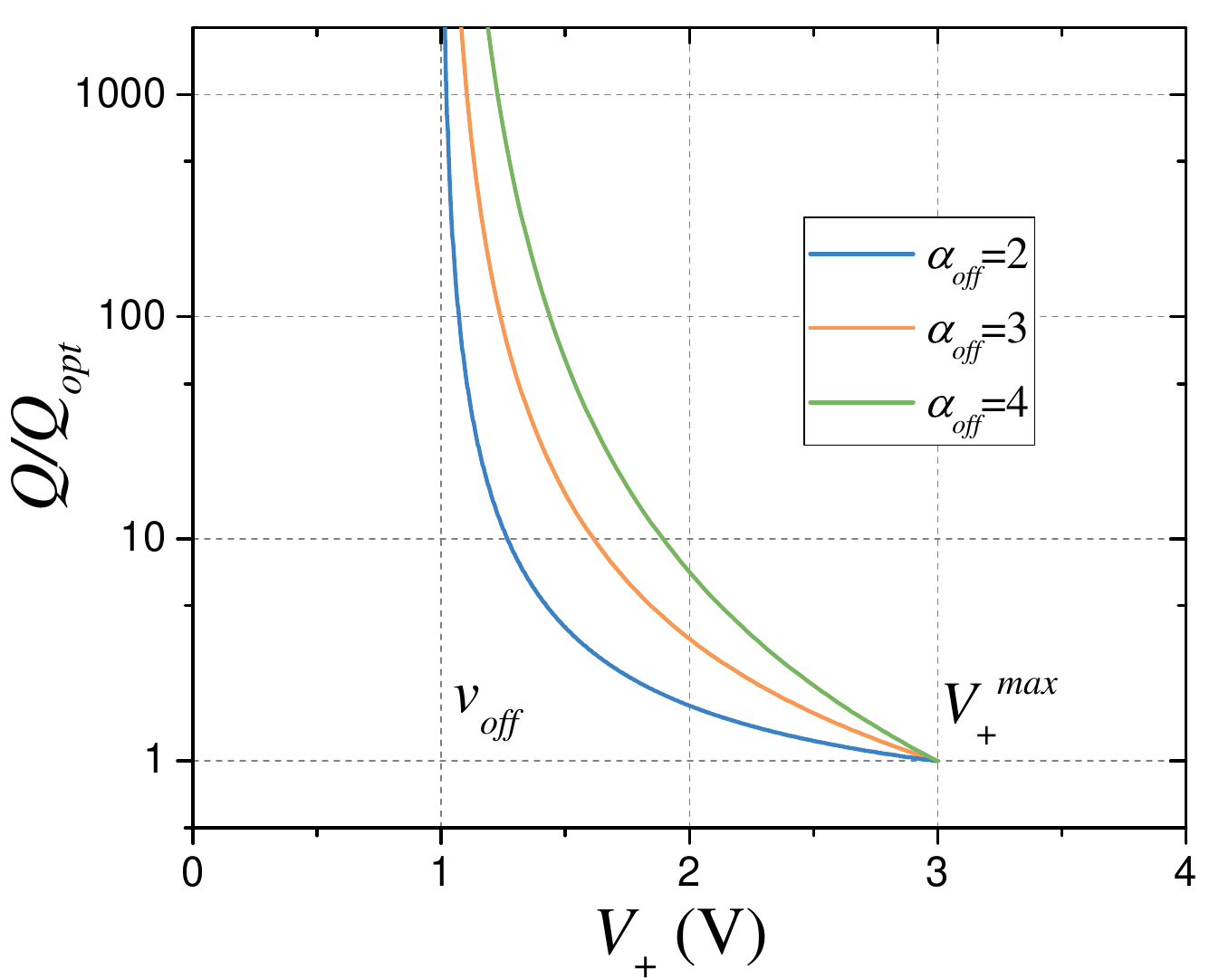}
    \caption{Energy dissipation relative to the optimal dissipation found for (a) $\alpha_{off/on}< 2$ and (b) $\alpha_{off/on}\geq 2$. These graphs were obtained using $v_{off}=1$~V, $v_{on}=-1$~V,  $\alpha_{on}=\alpha_{off}$, $V_{+}^{max}=3$~V, $V_{-}^{max}=-3$~V, and $V_-=-V_+$.}
    \label{fig:6}
\end{figure}

To illustrate the strategy outlined in Sect.~\ref{sec:3b}, Fig.~\ref{fig:6} displays the programming energy $Q$ versus the pulse amplitude $V_+$, relative to the optimal value $Q_{opt}$. This figure was generated for a device with symmetric kinetics, considering each value of the parameters $\alpha_{off}$ from the set $\{0.5,1,1.5\}$ in panel (a) and from the set $\{2,3,4\}$ in panel (b). In the present case, $Q/Q_{opt}$ remains constant regardless of the details of the trajectory, since the $I/T$ factor in Eq.~(\ref{eq:Qfinal}) is the same for the same trajectory $\bar{x}(t)$.

\begin{figure}[b]
     \includegraphics[width=0.35\columnwidth]{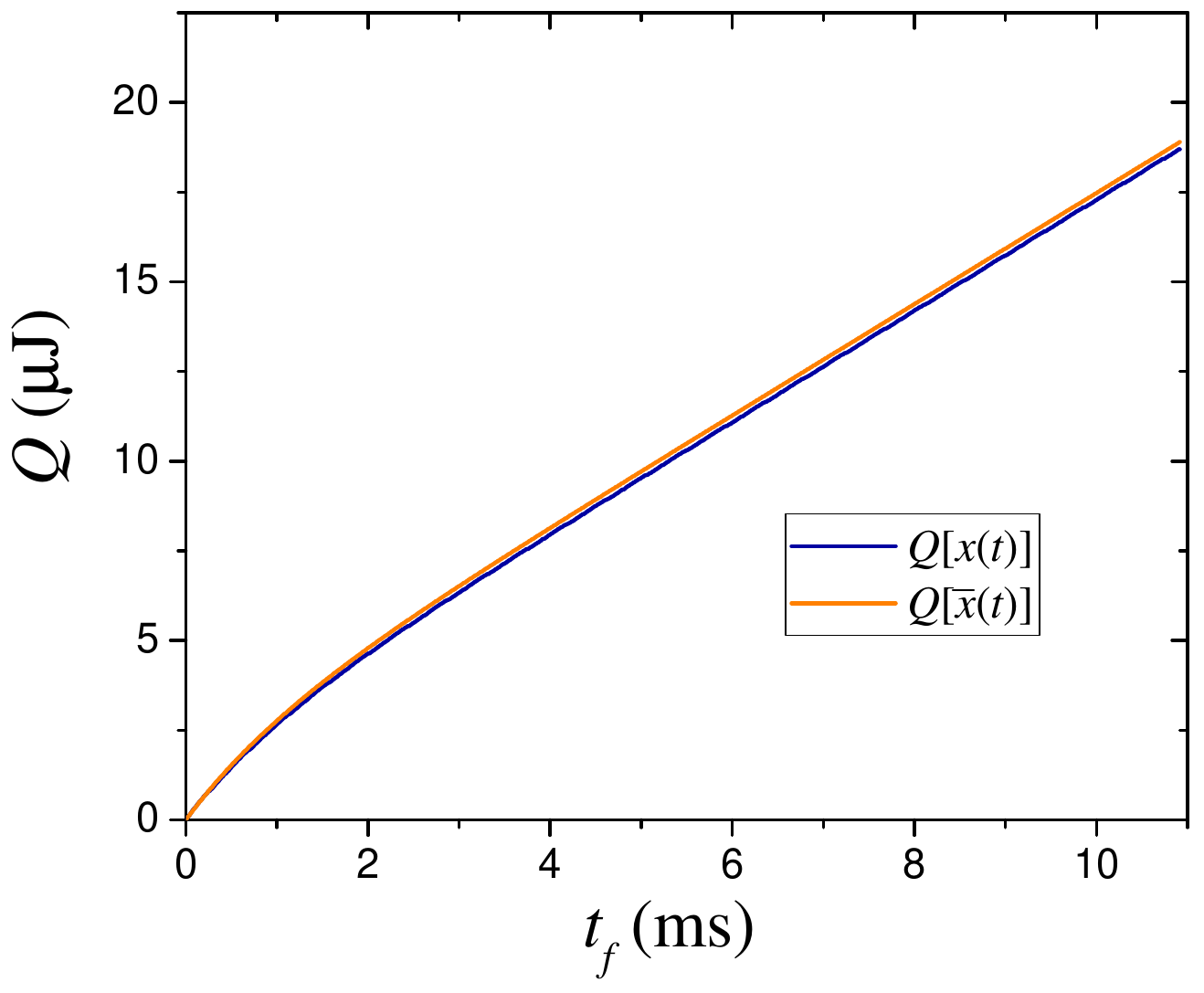}
    \hfill
    \begin{minipage}[b]{0.6\textwidth}
    \caption{Comparison between the Joule losses, expected to occur through the device in the programming scenario from Fig.~\ref{fig:5}, as determined from the exact trajectory $x(t)$, acquired from the numerical integration of the VTEAM model, and from the respective time-averaged solution $\bar{x}(t)$ on the basis of the proposed formula (\ref{eq:Qfinal}). 
    Here $G_{min}$ and $G_{max}$ were respectively taken equal  to $10^{-5}$~S and to $10^{-3}$~S.}\vspace{1cm}
    \label{fig:7}
    \end{minipage}
\end{figure}

For the simulation scenario illustrated in Fig.~\ref{fig:5}, Fig.~\ref{fig:7} presents a comparison between the Joule losses in the device, as calculated on the basis of the time-averaged trajectory $\bar{x}(t)$, utilizing our theoretical formula (\ref{eq:Qfinal}), and on the precise trajectory $x(t)$, employing equations~(\ref{eq:VTEAM_RESET})-(\ref{eq:VTEAM_SET}), respectively.

\section{Discussion} \label{sec:5}

Although many experiments on memristive devices have focused on the precision of state programming, another fundamental aspect, namely its energy efficiency, has attracted far less experimental attention~\cite{fleck2016energy}.
In this study, we have introduced a method to optimize the energy cost for programming a non-volatile resistive switching memory device using its dynamical attractor state~\cite{Pershin_2019}. This state is reached through the application of a specific sequence of alternating pulses from any initial state, thus in the presence of a
fading memory effect, as described in \cite{ascoli2016}. 
It has been found that \textcolor{black}{
the particular pulse sequence, that minimizes the programming
energy cost, depends upon several model parameters.}

\subsection{Programming sequence}

Figure~\ref{fig:3} illustrates an example in which the exponents $\alpha_{off}$ and $\alpha_{on}$ in the 
state evolution function $f(x,V)$ (see Eqs. (\ref{eq:VTEAM_RESET}) and (\ref{eq:VTEAM_SET})) are greater than or equal to 2. In this context, which assumes the VTEAM model is tailored to a device with asymmetric SET and RESET switching kinetics, the optimal strategy involves using pulses with the maximum potential amplitude ($|\hat{V}_\pm|=3$~V). 

Conversely, Fig.~\ref{fig:5} details a different case where the VTEAM model is adapted for a device with symmetric SET and RESET switching kinetics. Here, both $\alpha_{off}$ and $\alpha_{on}$ are less than 2, and the conditions $V_+^{max}>2\,v_{off}/(2-\alpha_{off})$ and $|V_-^{max}|>2|v_{on}|/(2-\alpha_{on})$ are satisfied. From Eq.~(\ref{eq:amplitudes+}) ((\ref{eq:amplitudes-})), it follows that an intermediate positive (negative) voltage level, specifically $\hat{V}_+=2 v_{off}/(2-\alpha_{off})=2$~V ($\hat{V}_-=2 v_{on}/(2-\alpha_{on})=-2$~V), should be used for the RESET and SET pulse heights, respectively, to achieve the most energy-efficient programming operation when applying the periodic voltage pulse train across the device. 

Figure~\ref{fig:4} was generated using the same model parameters as Fig.~\ref{fig:3}, except for $\epsilon$ and $x_a$, which are larger and smaller, respectively, than in the simulation that produced Fig.~\ref{fig:3}. By comparing Figs.~\ref{fig:3} and~\ref{fig:4}, it can be observed that increasing $\epsilon$ from $0.02$ to $0.1$ results in more pronounced oscillations in the state variable $x(t)$.

\subsection{Joule losses}

Figure~\ref{fig:6} illustrates the Joule losses predicted by our theoretical model as a function of the amplitude of the RESET pulse. In Fig.~\ref{fig:6}(a), a scenario is presented where $\alpha_{off} < 2$, suggesting a parameter-based RESET pulse configuration for optimal energy efficiency. In this case, the curve $Q/Q_{opt}$ versus $V_+$ has a minimum at $V_+=2v_{off}/(2-\alpha)$ for any value $\alpha$. 
However, \textcolor{black}{
the abscissa of the minimum of this curve} falls within (outside) the range $(v_{off},V_+^{max})$ for $\alpha_{off}\in(0,4/3)$ ($\alpha_{off}\in(4/3,2)$). In the latter situation, $\hat{V}_+$ should be set to the upper limit $V_+^{max}$ within the permissible range. 

In contrast, Fig.~\ref{fig:6}(b) presents the scenario where $\alpha_{off} \geq 2$, showing that the lowest energy cost is obtained when the maximum positive pulse is used. Here, regardless of the specific value assigned to $\alpha_{off}$, the energy $Q$ dissipated in the device, compared to its optimal counterpart $Q_{opt}$, consistently reduces to 1 when $V_+=V_+^{max}$. 

\textcolor{black}{Taking into account the properties of the curves} in Fig.~\ref{fig:6} can result in substantial energy savings during the critical and essential task of programming resistances in memristive devices.

\vspace{-0.5cm}
\textcolor{black}{
\subsection{Endurance and device-to-device variability effects}
This study assumes that the endurance is sufficient to perform programming via the dynamical attractor states throughout the device lifetime. In fact, a sequence of short pulses separated by zero voltage intervals is expected to cause less damage to the device than a single long pulse that produces the same change in $x$, because the long pulse deposits energy within a shorter time window. From a theoretical standpoint, however, determining the effects of device endurance limitations on the
proposed dynamic attractor state-based programming scheme is a difficult task. A study of
the impact of device-to-device variability on the accuracy of our programming paradigm can be found
in the Supplementary Information file.
}

\section*{Conclusion}

In conclusion, the dynamical attractor states of memristive devices offer an alternative \textcolor{black}{unsupervised} approach to their programming. This research has investigated methods to reduce energy consumption within this framework \textcolor{black}{focusing on devices described by} the VTEAM model. \textcolor{black}{Our approach can be extended to other device models and situations, including programming using current pulses. Moreover, a range of practical device limitations—such as variability, zero-input resistance drift, and environmental influences—can be incorporated into future modeling of dynamical attractors.}

The optimal pulse sequence depends on the critical parameters $\alpha_{off}$ and $\alpha_{on}$ \textcolor{black}{of the model}, which play a key role in the device's switching kinetics. When the former (latter) parameter is 2 or higher, the optimal energy-efficient strategy \textcolor{black}{envisages the selection of} the highest possible magnitude for RESET (SET) pulses \textcolor{black}{(fast switching)}. On the other hand, if $\alpha_{off}$ ($\alpha_{on}$) is less than 2, achieving maximum energy efficiency requires setting the RESET (SET) pulse magnitude to the lesser of a certain intermediate voltage magnitude \textcolor{black}{(slower switching)} and the highest possible voltage magnitude (taking into account the polarity).

\textcolor{black}{The results of this study} demonstrate that the choice of optimized pulse sequences can significantly decrease energy dissipation in memristive devices during programming \textcolor{black}{using their dynamical attractors}. This theoretical method has also been utilized to explore the reduction of energy dissipation in memristive devices under fast programming schemes, which could be beneficial for real-time edge computing applications.
\textcolor{black}{We hope our theoretical findings will motivate future experimental investigations aimed at evaluating the benefits of the proposed approach to program a non-volatile memristor device.}

%

\section*{Acknowledgement}
YVP was supported by the NSF grant EFRI-2318139.

\providecommand{\newblock}{}



\section*{Supplementary material}

# Supplementary information for the manuscript "Minimizing energy dissipation during programming of resistive switching memory devices using their dynamical attractor states"

Valeriy A. Slipko, Alon Ascoli, Fernando Corinto and Yuriy V. Pershin

## 1 Device variability effects

Let us consider how the variability in device parameters affects the position of the dynamical attractor $x_a$. According to the theory in the main text, the location of the dynamical attractor for the time-averaged dynamics is given by

$$x_a = \frac{A}{A-B}, \tag{S.1}$$

where

$$A = k_{off}\left(\frac{V_+}{v_{off}} - 1\right)^{\alpha_{off}} \frac{\tau_+}{T} \tag{S.2}$$

and

$$B = k_{on}\left(\frac{V_-}{v_{on}} - 1\right)^{\alpha_{on}} \frac{\tau_-}{T}, \tag{S.3}$$

with $\tau_+$ and $\tau_-$ calculated according to Eqs. (18) and (19) from the main text.

As a study example, we consider an ensemble of 50 000 memristors affected by device-to-device variability. To model this non-ideality, six VTEAM parameters—$k_{off}$, $k_{on}$, $v_{off}$, $v_{on}$, $\alpha_{off}$, and $\alpha_{on}$—are allowed to take random values, either in pairs or all together, sampled from Gaussian distributions centered at their nominal values with a given spread. Each random draw of the parameters defines one statistical realization. For each realization, the actual location of the dynamical attractor is computed analytically using Eq. (S.1) together with Eqs. (S.2) and (S.3), while keeping $\tau_+$ and $\tau_-$ to the values they hold under the nominal parameter values. This approach avoids running 50 000 time-domain simulations of the VTEAM model.

The probability that the attractor lies in the interval $[x, x + \Delta x]$ is $P_a(x, x + \Delta x) = p_a(x)\Delta x$, where $p_a(x)$ is the corresponding probability density function.

Following the example in Fig. 2 of the main text, the programming pulse sequence is chosen such that, ideally, $x_a = 0.6$. The blue distributions in Fig. S.1(a) and (b) show $p_a(x)$ when all six parameters are varied with standard deviations of 1% and 5%, respectively. The same panels of Fig. S.1 also report three additional cases where only $k_{off}, k_{on}$ (yellow), $v_{off}, v_{on}$ (red), or $\alpha_{off}, \alpha_{on}$ (violet) are randomized, while the remaining parameters are fixed at their nominal values.

Comparison between Figs. S.1(a) and (b) shows that increasing the spread of the parameters leads to a broader distribution in the location of the attractor. When all six parameters vary, the resulting distributions in the location of the attractor are well approximated by Gaussians with standard deviations $\sigma =$

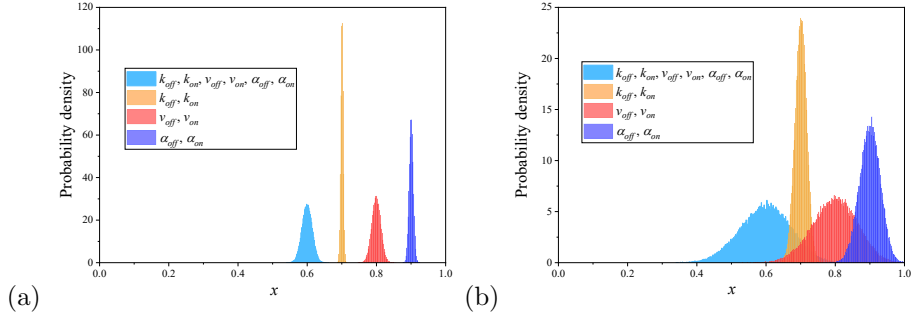

(a)  (b)

Figure S.1: Histograms of probability density function $p_a(x)$ for a periodically driven VTEAM memristor. The nominal model parameters are $k_{off} = 10^3$ s$^{-1}$, $k_{on} = -10^3$ s$^{-1}$, $v_{off} = 1$ V, $v_{on} = -1$ V, $\alpha_{off} = 3$, and $\alpha_{on} = 2$. The programming pulse sequence is defined by $\hat{V}_+ = 3$ V, $\hat{V}_- = -3$ V, $\tau_+ = 6.3$ µs, $\tau_- = 8.3$ µs, and $T = 34.6$ µs. The six model parameters are assumed to vary either in pairs—$(k_{off}, k_{on})$ (yellow), $(v_{off}, v_{on})$ (red), and $(\alpha_{off}, \alpha_{on})$ (violet)—or all together (blue), according to Gaussian distributions centered at their nominal values, with standard deviations of 1% in (a) and 5% in (b). For clarity, the distributions obtained by varying parameter pairs are shifted horizontally relative to one another. Each distribution is computed from the closed-form expression (S.1) using 50 000 statistical realizations.

0.015 and $\sigma = 0.072$ in plots (a) and (b) of Fig. S.1, respectively. Comparing the effects of spreads in the three parameter pairs $(k_{off}, k_{on})$, $(v_{off}, v_{on})$, $(\alpha_{off}, \alpha_{on})$, the variability in $(v_{off}, v_{on})$ is found to have the strongest impact.

Finally, Fig. S.2 presents a comparison between the blue distributions from Fig. S.1 and those corresponding to the scenario, where the memristor, whose state is initially equal to $x(0) = 0.1$, is programmed by a single square RESET voltage pulse designed to raise the memristor state to 0.6. Using Eq. (20) in [A1], the pulse duration is selected according to

$$T_{off} = \frac{x_{off} - x_{on}}{k_{off}(V_0/v_{off} - 1)^{\alpha_{off}}} \ln \frac{x_{off} - x_i}{x_{off} - x_f}, \qquad (S.4)$$

with $x_{off} = 1$, $x_{on} = 0$, $x_i = x(0)$, $x_f = x_a$, and $V_0 = \hat{V}_+$. In both panels of Fig. S.2, the two distributions are very close to each other, indicating that programming via alternating-polarity pulse sequences does not introduce additional significant errors compared to single-pulse, initial-condition-dependent programming.

# References

[S1] Slipko V A, Ascoli A, Corinto F and Pershin Y V 2025 *arXiv preprint arXiv:2508.15620*

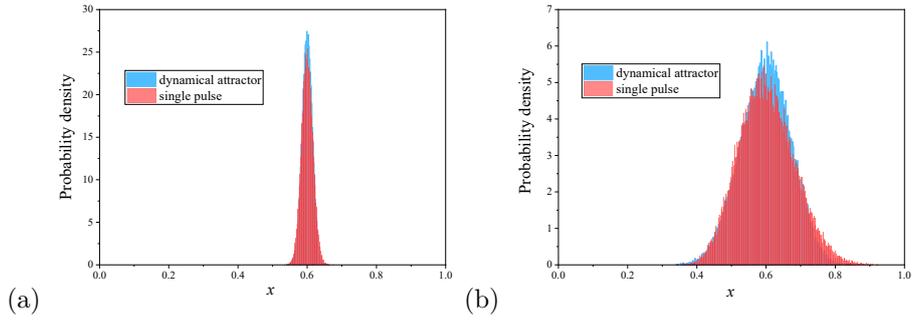

Figure S.2: Comparison between the probability density function of the dynamical attractor, obtained from 50 000 statistical realizations with all model parameters allowed to vary randomly (blue distributions in Fig. S.1(a) and (b) for 1% and 5% Gaussian spreads, respectively), and the probability density function for the state of the memristor, programmed by a single square voltage pulse of $V_0 = 3$ V amplitude and $T_{off} = 101.4$ µs width. The latter is computed from 50 000 numerical realizations of the VTEAM model, using the same parameter values as in Fig. S.1 and their statistical distributions, $x(0) = 0.1$, and the target value $x = 0.6$.

3




\begin{thebibliography}{10}
\expandafter\ifx\csname url\endcsname\relax
  \def\url#1{{\tt #1}}\fi
\expandafter\ifx\csname urlprefix\endcsname\relax\def\urlprefix{URL }\fi
\providecommand{\eprint}[2][]{\url{#2}}

\bibitem{ielmini_waser_2016}
Ielmini D and Waser R 2016 {\em Resistive Switching: From Fundamentals of Nanoionic Redox Processes to Memristive Device Applications\/} (Wiley-VCH)

\bibitem{5405039}
Pershin Y~V and Di~Ventra M 2010 {\em IEEE Trans. Circuits Syst. I\/} {\bf 57} 1857--1864

\bibitem{pershin2010experimental}
Pershin Y~V and Di~Ventra M 2010 {\em Neural networks\/} {\bf 23} 881--886

\bibitem{9618724}
Kang S~M, Choi D, Eshraghian J~K, Zhou P, Kim J, Kong B~S, Zhu X, Demirkol A~S, Ascoli A, Tetzlaff R, Lu W~D and Chua L~O 2021 {\em IEEE Trans. Circuits Syst. I\/} {\bf 68} 4837--4850

\bibitem{ascoli2020}
Ascoli A, Tetzlaff R, Kang S and Chua L 2020 {\em IEEE Trans. Circuits Syst. I\/} {\bf 67} 2753 -- 2766

\bibitem{Strachan18a}
Hu M, Graves C~E, Li C, Li Y, Ge N, Montgomery E, Davila N, Jiang H, Williams R~S, Yang J~J, Xia Q and Strachan J~P 2018 {\em Advanced Materials\/} {\bf 30} 1705914

\bibitem{Teuscher17a}
Woods W and Teuscher C 2017 Approximate vector matrix multiplication implementations for neuromorphic applications using memristive crossbars {\em 2017 IEEE/ACM International Symposium on Nanoscale Architectures (NANOARCH)\/} pp 103--108

\bibitem{Amirsoleimani20a}
Amirsoleimani A, Alibart F, Yon V, Xu J, Pazhouhandeh M~R, Ecoffey S, Beilliard Y, Genov R and Drouin D 2020 {\em Advanced Intelligent Systems\/} {\bf 2} 2000115

\bibitem{Pershin_2019}
Pershin Y~V and Slipko V~A 2019 {\em Europhysics Letters\/} {\bf 125} 20002

\bibitem{pershin_2019_bif_anal_TaO_mod}
Pershin Y~V and Slipko V~A 2019 {\em Journal of Physics D: Applied Physics\/} {\bf 52} 505304

\bibitem{Slipko19a}
Slipko V~A and Pershin Y~V 2019 {\em Physica E: Low-dimensional Systems and Nanostructures\/} {\bf 114} 113561

\bibitem{asc_front_electr_2023}
Ascoli A, Schmitt N, Messaris I, Demirkol A, Tetzlaff R and Chua L 2023 {\em Frontiers in Electronic Materials\/} {\bf 23} 1228899

\bibitem{schmitt2024theoretico}
Schmitt N, Ascoli A, Messaris I, Demirkol A, Menzel S, Rana V, Tetzlaff R and Chua L 2024 {\em Frontiers in Nanotechnology\/} {\bf 6} 1301320

\bibitem{messaris2023}
Messaris I, Demirkol A, Ascoli A and Tetzlaff R 2023 {\em IEEE Trans. Circuits Syst. I\/} {\bf 70} 566--578

\bibitem{ascoli2016}
Ascoli A, Tetzlaff R, Chua L, Strachan J and Williams R 2016 {\em IEEE Trans. Circuits Syst. I\/} {\bf 63} 389--400

\bibitem{slipko2024reduction}
Slipko V~A and Pershin Y~V 2025 {\em IEEE Transactions on Nanotechnology\/} {\bf 24} 8--16

\bibitem{astin2025low}
Astin N and Pershin Y~V 2025 {\em arXiv preprint arXiv:2507.18487\/}

\bibitem{slipko2025low}
Slipko V~A, Ascoli A, Corinto F and Pershin Y~V 2025 {\em arXiv preprint arXiv:2508.15620\/}

\bibitem{Kvatinsky2013}
Kvatinsky S, Ramadan M, Friedman E~G and Kolodny A 2015 {\em IEEE Trans. Circuits Syst. II\/} {\bf 62} 786--790

\bibitem{chua76a}
Chua L~O and Kang S~M 1976 {\em Proceedings of {IEEE}\/} {\bf 64} 209--223

\bibitem{kim2020experimental}
Kim J, Pershin Y~V, Yin M, Datta T and Di~Ventra M 2020 {\em Advanced Electronic Materials\/} {\bf 6} 2000010

\bibitem{chua71a}
Chua L~O 1971 {\em {IEEE} Transactions on Circuit Theory\/} {\bf 18} 507--519

\bibitem{arfken2011mathematical}
Arfken G~B, Weber H~J and Harris F~E 2011 {\em Mathematical methods for physicists: {A} comprehensive guide\/} (Academic press)

\bibitem{fleck2016energy}
Fleck K, B{\"o}ttger U, Waser R, Aslam N, Hoffmann-Eifert S and Menzel S 2016 Energy dissipation during pulsed switching of strontium-titanate based resistive switching memory devices {\em 2016 46th European Solid-State Device Research Conference (ESSDERC)\/} (IEEE) pp 160--163

\end{thebibliography}
\end{document}